\documentclass[aps,prx,twocolumn,superscriptaddress, nofootinbib,  longbibliography, floatfix]{revtex4-1}
\usepackage{graphicx} % needed for figs
\usepackage{dcolumn}  % needed for some tables
\usepackage{bm}    % for math
\usepackage{amssymb}  % for math
\usepackage{amsfonts, amsmath, amssymb, mathtools}
\usepackage{amsmath}

\DeclareMathOperator{\Tr}{Tr}
\usepackage{lipsum}
\usepackage[normalem]{ulem}
\usepackage{commath}
\usepackage{xcolor}  
\usepackage{placeins}
\usepackage{scalerel}[2016-12-29]

\newcolumntype{P}[1]{>{\centering\arraybackslash}p{#1}}
\usepackage[binary-units=true]{siunitx}
\usepackage{cleveref}
\RequirePackage{textcase}
\DeclareUnicodeCharacter{2009}{\,} 
\renewcommand{\selectlanguage}[1]{}

\begin{document}

\newcommand{\todo}[1]{{\color[rgb]{1.0,0.0,0.0}#1}} % gaps to be filled in red

\newcommand{\blue}[1]{\textcolor{blue}{#1}}
\newcommand{\green}[1]{\textcolor{green}{#1}}
\newcommand{\yg}[1]{\textcolor{purple}{#1}}

\raggedbottom

\renewcommand{\thesubsection}{\thesection.\arabic{subsection}}
\renewcommand{\thesubsubsection}{\thesubsection.\arabic{subsubsection}}

\title{Demonstrating efficient and robust bosonic state reconstruction\\via optimized excitation counting}

\author{Tanjung Krisnanda}
\thanks{These authors contributed equally to this article. The order of author names can be re-arranged in individual CVs.}
\author{Clara Yun Fontaine}
\thanks{These authors contributed equally to this article. The order of author names can be re-arranged in individual CVs.}
\author{Adrian Copetudo}
\thanks{These authors contributed equally to this article. The order of author names can be re-arranged in individual CVs.}
\author{Pengtao Song}
\thanks{These authors contributed equally to this article. The order of author names can be re-arranged in individual CVs.}
\affiliation{Centre for Quantum Technologies, National University of Singapore, Singapore 117543, Singapore}
\author{\\Kai Xiang Lee}
\affiliation{School of Physical and Mathematical
Sciences, Nanyang Technological University, Singapore 637371, Singapore}
\author{Ni-Ni Huang}
\affiliation{Centre for Quantum Technologies, National University of Singapore, Singapore 117543, Singapore}
\author{Fernando Valadares}
\affiliation{Centre for Quantum Technologies, National University of Singapore, Singapore 117543, Singapore}
\author{Timothy C. H. Liew}
\affiliation{School of Physical and Mathematical
Sciences, Nanyang Technological University, Singapore 637371, Singapore}
\author{Yvonne Y. Gao}
\email[Corresponding author: ]{yvonne.gao@nus.edu.sg}
\affiliation{Centre for Quantum Technologies, National University of Singapore, Singapore 117543, Singapore}
\affiliation{Department of Physics,
National University of Singapore, Singapore 117542, Singapore}
\date{\today}

\begin{abstract}
Quantum state reconstruction is an essential element in quantum information processing. However, efficient and reliable reconstruction of non-trivial quantum states in the presence of hardware imperfections can be challenging. This task is particularly demanding for high-dimensional states encoded in continuous-variable (CV) systems, where a large number of grid-based measurements are often used to adequately sample relevant regions of phase space. In this work, we introduce an efficient and robust technique of Optimized Reconstruction with Excitation Number Sampling (ORENS) based on the idea of generalized Q-function. We use a standard bosonic circuit quantum electrodynamics (cQED) setup to experimentally demonstrate effective state reconstruction using the theoretically minimum number of measurements. 
Our investigation highlights that ORENS is naturally free of parasitic system dynamics and resilient to decoherence effects in the hardware, enabling it to outperform the conventional reconstruction techniques in cQED such as Wigner tomography. Finally, ORENS relies only on the ability to accurately measure the excitation number of a given CV state, making it a versatile and accessible tool for a wide range of CV platforms and readily scalable to multimode systems. Thus, our work provides a crucial and valuable primitive for practical quantum information processing using bosonic modes.
\end{abstract}

\maketitle

%%%%%%%%%%%%%%%%%%%%%%%%%%%%%%%%%%%%%%%%%%%%%%%%%%%
%\section{Introduction}
%%%%%%%%%%%%%%%%%%%%%%%%%%%%%%%%%%%%%%%%%%%%%%%%%%%

\section{\label{sec:level1}INTRODUCTION}
Continuous-variable (CV) quantum systems offer the rich and versatile dynamics of a large Hilbert space~\cite{copetudo_shaping_2024, weedbrook_gaussian_2012, pan_continuous-variable_2023, joshi_quantum_2021}, with applications ranging across quantum computation~\cite{campagne-ibarcq_quantum_2020,gao_entanglement_2019}, metrology~\cite{wang_heisenberg-limited_2019}, and simulation~\cite{liu_twofold_2017}. To take full advantage of these systems, it is essential to develop techniques to accurately characterize the properties, interactions, and evolutions of their quantum states. However, reconstructing the density matrix of an arbitrary CV state in a large Hilbert space is a challenging task. Not only are many measurement observables needed to capture features spread across the large phase space, but experimentally, the observables must often be mapped to an auxiliary element (e.g. a qubit) via non-ideal and error-prone operations to extract the relevant measurement outcomes. 

Current techniques used in CV quantum information platforms include the Wigner~\cite{lutterbach_method_1997, bertet_direct_2002}, Husimi Q~\cite{husimi1940some}, and generalized Q~\cite{opatrny_density-matrix_1997,kirchmair_observation_2013} function measurements. They rely on oversampling the phase space features of the CV state to achieve high reconstruction accuracy. However, as these quantum systems advance in both scale and complexity, it is necessary to design more optimal reconstruction techniques that use the fewest measurements of observables that are resilient against experimental errors. While many different approaches to tackling this scalability challenge have been theoretically proposed or experimentally demonstrated~\cite{landon-cardinal_quantitative_2018, he_efficient_2023, chakram_multimode_2022,shen_optimized_2016, wang_schrodinger_2016}, these strategies often come at the cost of versatility, measurement quality, engineering convenience, and optimization complexity.

In this work, we present a method to robustly and efficiently reconstruct arbitrary bosonic states with the optimized fewest measurements of excitation number. Our technique, named Optimized Reconstruction with Excitation Number Sampling (ORENS), augments the classic idea of generalized Q function with a carefully designed optimization procedure to allow effective reconstruction of arbitrary CV states with minimal measurements. 
For instance, compared to an earlier work, Ref.~\cite{kirchmair_observation_2013}, which reconstructs states within the truncation dimension $D=8$, ORENS offers a 50-fold reduction in the number of measurement observables without compromising reconstruction accuracy. The technique can be readily implemented across CV platforms, where these excitations take the form of optical photons~\cite{calkins_high_2013}, microwave photons~\cite{brune_quantum_1996, guerlin_progressive_2007, schuster_resolving_2007, wang_decoherence_2009}, and motional phonons of trapped ions~\cite{leibfried_experimental_1996, an_experimental_2015, lo_spinmotion_2015}. We experimentally demonstrate the efficacy of ORENS in a cQED platform to showcase its performance for high-dimensional CV states, even under severe decoherence. We show that ORENS outperforms the state-of-the-art Wigner reconstruction technique~\cite{lutterbach_method_1997, bertet_direct_2002}, owing to its inherent robustness against both coherent and incoherent errors. Our contribution to bosonic state reconstruction, a key research pillar in CV applications, will reinforce the development and analysis of more complex bosonic states and dynamics across different physical devices. 

\begin{figure*}
\centering
\includegraphics[width=1.0\textwidth]{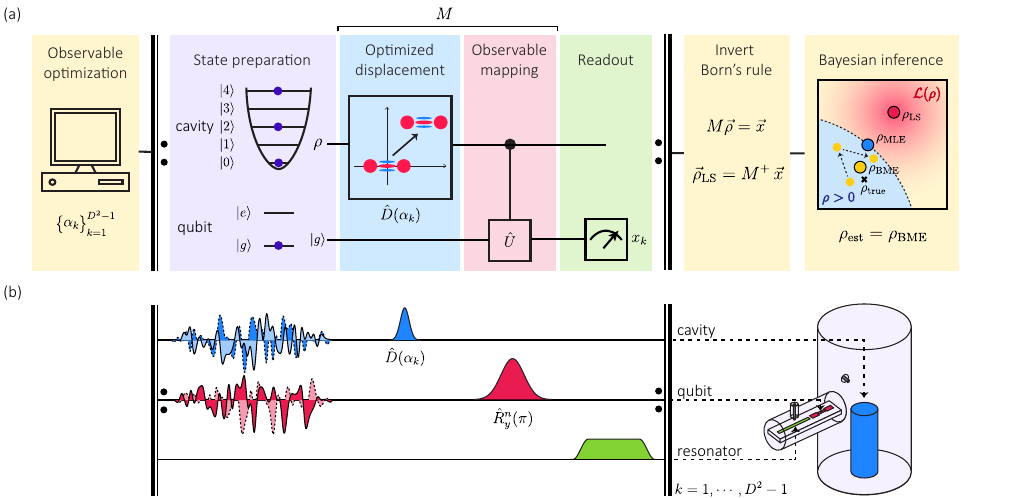}

\caption{{\bf An overview of bosonic state reconstruction and its translation to ORENS in cQED.} 
(a) To reconstruct an arbitrary CV state $\rho$ of dimension $D$, measurements of at least $D^2-1 \coloneqq N^*_{obs}$ independent real parameters must be made to be informationally complete. In bosonic systems, this is typically achieved through displacements, $\hat D(\alpha_k)$, to $N^*_{obs}$ distinct points in phase space which effectively changes the measurement observable. These displacement points, $\{\alpha_k\}$ are classically optimized before the experiment for a specific truncation dimension, without any knowledge of the state itself. A Bayesian inference algorithm is applied to compute the optimal estimator $\rho_\text{BME}$ based on the resulting measurement outcomes~\cite{lukens_practical_2020}. All relevant optimization procedures and data processing codes are available on GitHub. (b) A prototypical bosonic cQED device consists of a storage cavity (blue), an ancillary qubit (red), and a readout resonator (green). State preparation is implemented with numerically optimized pulses played to the cavity and qubit. 
Displacements on the cavity state are implemented by driving the cavity at its resonance frequency using a microwave pulse with a Gaussian envelope.
The excitation number sampling for ORENS is enacted by a $\pi$-pulse on the qubit conditioned on a specific photon number, $n$, in the bosonic mode. The optimal choice of $n$, is made before the experiment based only on the truncation dimension. The measurement outcome is extracted with the standard dispersive readout of the resonator.} 
\label{fig: Fig1}
\end{figure*}

\section{\label{sec:level1}ORENS PROTOCOL}
Conceptually, reconstructing an arbitrary quantum state $\rho$ consists of measuring identically prepared copies along many different bases. To accurately obtain information about the state, these bases must be informationally complete and the measurements have to be resilient against errors. For CV systems, when the state does not extend beyond a certain dimension $D$, its Hilbert space can be truncated~\cite{RevModPhys.81.299, deleglise2008reconstruction}. As such, only $D^2-1 \coloneqq N^*_{obs}$ independent real parameters must be obtained from measurements for informational completeness, setting the optimal number of measurements for efficient reconstruction~\cite{lukens_practical_2020}. To achieve independence between measurements for CV systems, distinct displacement transformations $\hat{D}(\alpha) = \exp\left(\alpha \hat{c}^\dagger + \alpha^* \hat{c}\right)$ can be conveniently applied on $\rho$ to sample different regions of phase space~\cite{RevModPhys.81.299}. 
% Alternatively, the measurement observable sampled at a particular point in phase-space can be changed. 
Upon choosing a base observable and a set of displacements, the set of measurements is written as a matrix $M$, and the measurement process is described with Born's rule as $\vec x = M \vec \rho$, where $\vec x$ is the measurement outcomes and $\vec \rho$ is the vectorized density matrix. Born's rule can be inverted to find the least-squares estimator $\vec \rho_\text{LS}=M^{+}\vec x$, where $M^+$ is the Moore-Penrose pseudo-inverse of $M$ (Appendix~\ref{apx_sub:linear-inversion}), which is constrained to be physical to realize the final estimator $\rho_\text{est}$ (Appendix~\ref{apx_sub:bayesian}). An overview of the key bosonic state reconstruction stages is illustrated in Fig.~\ref{fig: Fig1}a. 

Ideally, measurements of the minimal $N^*_{obs}$ independent real parameters enable perfect reconstruction of $\rho$. However, with experimental imperfections, the accuracy of the estimated state critically depends on the choice of $M$, which amplifies measurement errors to varying degrees upon estimation. The robustness to error is characterized by the condition number (CN) of the measurement matrix $M$, where a CN of 1 corresponds to the absence of error amplification and grants the optimal reconstruction~\cite{bhatia_matrix_1997}. 

Our proposed method, ORENS, leverages the sampling of excitation number across phase space, which is not only a readily accessible measurement observable in many CV platforms but also has built-in robustness against both dephasing and non-ideal coherent dynamics. Prior to the experiment, a minimal set of $N^*_{obs}$ measurement observables is classically optimized using a standard gradient descent method for a given truncation dimension $D$. This provides a set of optimal displacement points and excitation numbers that effectively minimizes the CN (Appendix~\ref{apx_sub:optimizing-set-measurements}) and ensures the accuracy of the reconstruction. This classical optimization process only needs to be done once and the resulting set of displacements can be used to reconstruct any arbitrary CV states within the same truncation dimension.

% With the minimal $N^*_{obs}$ excitation number measurements preceded by the optimized displacements in phase space that minimize the CN to state-of-the-art (Appendix~\ref{apx_sub:optimizing-set-measurements}), ORENS is capable of reliably reconstructing the density matrix of an arbitrary complex CV state. 

% \textcolor{blue}{Our proposed method, ORENS, leverages the sampling of excitation number across phase space to robustly and efficiently reconstruct the density matrix of an arbitrary CV state.}to state-of-the-art error resilience (Appendix~\ref{apx_sub:linear-inversion} and \ref{apx_sub:optimizing-set-measurements}). The inherent robustness against experimental errors -- decoherence and non-ideal coherent dynamics -- as well as the accessibility of excitation number sampling, makes ORENS an attractive reconstruction technique across many CV platforms.}

ORENS is conceptually based on the technique of generalized Q-function~\cite{kirchmair_observation_2013, opatrny_density-matrix_1997, lundeen_tomography_2009, zhang_mapping_2012}, where $Q_n(\alpha)=\text{Tr} \left( |n\rangle \langle n|\hat D(\alpha)^{\dagger}\rho \hat D(\alpha) \right)$. This is the generalization of the Husimi-Q function, $Q_0(\alpha) = \langle \alpha | \rho | \alpha \rangle$ to an arbitrary number of excitations $n$. Sampling higher $n$ overcomes the limitations of Husimi Q by boosting sensitivity to phase-space oscillations of $\rho$. This can be understood graphically by considering that the $Q_n$ function of a given state $\rho$ is the convolution of the Wigner function of $\rho$ with that of $|n\rangle$~\cite{smith_generalized_2006}. For the specific case of vacuum $n=0$, the Wigner of vacuum is a Gaussian distribution centered in the origin of the phase space and thus acts as a Gaussian filter on $\rho$, erasing fast phase-space oscillations and results in a strictly non-negative version of the Wigner function. However, for larger $n>0$, the features of $\rho$ are better preserved by $Q_n$, enabling robust reconstruction.

We validate the resilience and efficiency of ORENS by reconstructing arbitrary CV states in cQED, where excitation number is synonymous with photon number. We demonstrate that the excitation number measurement is inherently free of parasitic system dynamics and robust against decoherence. In our hardware, illustrated in Fig.~\ref{fig: Fig1}b, the CV states are stored in the electromagnetic field of superconducting LC resonators, realized as a high-Q 3D coaxial cavity machined out of high-purity (4N) aluminum. The states are prepared, transformed, and measured via the engineered dispersive interaction with an auxiliary qubit. The qubit is a standard transmon dispersively coupled to an on-chip readout resonator, and both elements are fabricated out of aluminum on a sapphire substrate  (Appendix~\ref{apx_sub:package_chip_fab}). The full Hamiltonian of this qubit-cavity system can be found in Appendix~\ref{apx_sub:hamiltonian}.
% \begin{equation}   
% \begin{aligned}
%     \frac{\mathbf{H}}{\hbar} & = \omega_q \hat{\mathbf{q}}^\dagger \hat{\mathbf{q}} 
%     + \frac{\alpha}{2} \hat{\mathbf{q}}^\dagger \hat{\mathbf{q}}^\dagger \hat{\mathbf{q}} \hat{\mathbf{q}} 
%     + \omega_c \hat{\mathbf{c}}^\dagger \hat{\mathbf{c}} 
%     - \chi \hat{\mathbf{q}}^\dagger \hat{\mathbf{q}} \hat{\mathbf{c}}^\dagger \hat{\mathbf{c}} \\
%     & + \frac{\Omega_d (t)}{2} \left( 
%     \hat{\mathbf{q}} e^{-i(\omega_d + \phi_d)} + \hat{\mathbf{q}}^\dagger e^{i(\omega_d + \phi_d)} \right),
% \end{aligned}
% \label{eq: Hamiltonian}
% \end{equation}
% where $\omega_q, \omega_c$ are the angular frequencies and $\hat{\mathbf{q}}^\dagger (\hat{\mathbf{q}})$ and $\hat{\mathbf{c}}^\dagger (\hat{\mathbf{c}})$ are the creation (annihilation) operators of the qubit and the cavity, respectively; $\alpha$ is the anharmonicity of the qubit, $\chi$ is the dispersive coupling between the cavity and the qubit, and $\Omega_d (t)$, $\omega_d$ and $\phi_d$ are the time-dependent amplitude, angular frequency, and phase of the drive, respectively. %In cQED, the excitation number in the cavity $c^\dagger c = n$ is synonymous with excitation number.

\section{\label{sec:level1}EXCITATION COUNTING IN CQED}

Extracting the excitation number of the cavity consists of conditionally exciting the qubit depending on the number of excitations in the cavity. This conditional excitation leverages the dispersive interaction between the cavity and qubit, which can be understood by looking at the system Hamiltonian in the rotating frame of the qubit drive,
\begin{equation}
    \frac{\hat H}{\hbar} = \frac{\Omega_d}{2} \hat \sigma_y + \Delta |e\rangle \langle e| - \chi \hat n |e\rangle \langle e|,
    \label{eq: rotation}
\end{equation}
where $\hat n$ is the excitation number operator of the cavity mode, $\chi$ is the dispersive coupling between the cavity and the qubit, $\hat \sigma_y$ is the $\hat y$ Pauli operator, $\Delta$ is the drive detuning from the qubit frequency, and $\Omega_d$ is the drive amplitude. The first term in Eq.~(\ref{eq: rotation}) yields a rotation of the qubit state around the $y$-axis of the qubit Bloch sphere, where the rotation angle depends on the drive amplitude $\Omega_d$ and the pulse length. A $\pi$-rotation corresponds to a complete flip of the qubit state. By using a long, spectrally-selective drive with duration denoted as $t_\pi$, such that  $t_\pi> 1 / \chi$, the individual shifts of the qubit frequency ($\omega_q - n \chi$) corresponding to $n$ excitations in the cavity, are resolved and can be individually addressed~\cite{schuster_resolving_2007}. Choosing a drive detuning $\Delta = \chi n$ enables robust mapping of the cavity's excitation number to the qubit:
\begin{equation}
    p_n= \text{Tr} \left( \rho |n\rangle \langle n| \right) \approx p_e(\Delta=\chi n),
\label{eqn:perfect_pn_coherent}
\end{equation}
where $p_n$ is the probability of the cavity having $n$ excitations, and $p_e$ is the probability of the qubit being in the excited state. More details can be found in Appendix~\ref{apx_sub:ens-mapping-selectivity}.

\begin{figure}[!]
\centering
\includegraphics{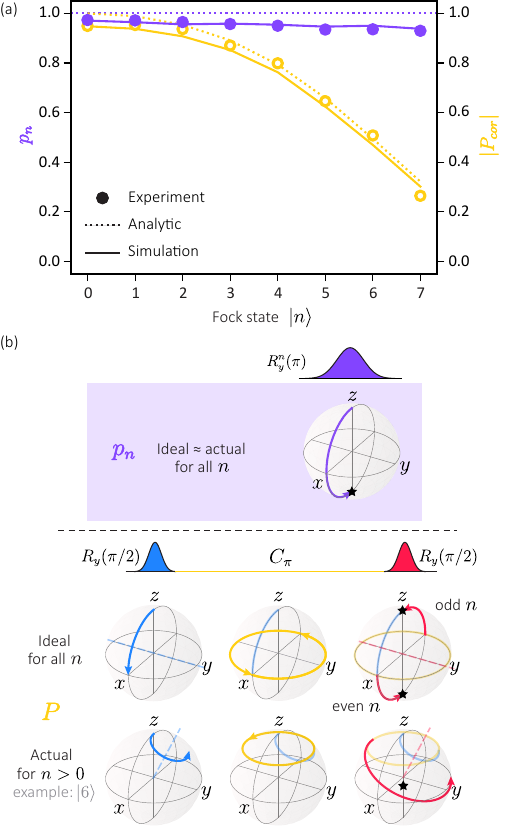}
\caption{{\bf Excitation number and parity mapping of Fock states in cQED.} 
(a) Measurement outcomes of excitation number $p_n$ (purple) and absolute value of corrected parity $|P_\text{cor}|$ (yellow) as a function of the Fock state $|n\rangle$ prepared in the cavity. Data is obtained with qubit $T_{\phi}$ of 15.3\,$\mu$s. They show good agreement with analytical (Eq.~(\ref{eqn:perfect_pn_coherent}) and Eq.~(\ref{eqn:imperfect_Pcor_coherent})) and simulated trends based on real device parameters. (b) Bloch spheres of the qubit state at each step of excitation number (above) and parity (below) mapping. The measurement outcomes are marked with black stars.
}
\label{fig: Fig2}
\end{figure}

By design, this mapping of the excitation number does not experience significant parasitic Hamiltonian dynamics, making it an excellent choice of measurement observable for state reconstruction. To verify this, we prepare a given Fock state $|n \rangle$ in the cavity using numerically optimized Gradient Ascent Pulse Engineering (GRAPE) pulses~\cite{heeres_implementing_2017}. We flip the state of the qubit with a $t_\pi = 1 \mu$s Gaussian pulse selective at a frequency $\omega_q - \chi n$ to map the excitation number corresponding to $|n \rangle$ to the state of the qubit, which is read out with a single-shot measurement. We repeat the experiment 1000 times and use the average outcomes to estimate $Q_n(0)$. The results, shown in Fig.~\ref{fig: Fig2}a, demonstrate effective mapping of the excitation number from the cavity to the qubit state with $p_n>0.93$. This is in excellent agreement with the simulated outcomes, where the effects of qubit decoherence, qubit thermal population, and finite readout discrimination are taken into account.

In contrast to excitation number measurement, the cavity parity measurement commonly used in the ubiquitous Wigner reconstruction is highly prone to coherent errors in the system. For the parity measurement, an $\pi/2$-rotation pulse brings the state of the qubit, originally in $|g\rangle$, to the equator of the Bloch sphere, $|+\rangle$, by evolving the qubit state under the Hamiltonian $\hat H= (\Omega/2) \hat \sigma_y$ for a time $\pi/(2\Omega)$. Then, a conditional phase gate, $\hat C_{\pi}=|g\rangle \langle g|\otimes \openone+|e\rangle \langle e|\otimes e^{i\pi \hat n}$, is implemented by waiting for a time $t_{w} = \pi/\chi$. As such, the state of the qubit acquires a relative phase depending on the photon number parity, with its resulting state being in $|+\rangle (|-\rangle)$ when the cavity contains an even (odd) number of excitations. A final $\pi/2$-rotation pulse maps $|+\rangle (|-\rangle)$ to $|e\rangle (|g\rangle)$. Despite this protocol being a good approximation to estimate the parity, the always-on dispersive interaction during the $\pi/2$ pulses imparts substantial error to the measurements, with the effective Hamiltonian being $\hat H= (\Omega/2) \hat \sigma_y - \chi \hat n |e\rangle \langle e|$. When the cavity has $n>0$ excitations, the qubit rotations happen along a slanted axis, given by $\hat r = (\Omega \hat y+\chi n \hat z)/\sqrt{\Omega^2+(\chi n)^2}$. As a result, the qubit state prematurely accumulates phase during the first $\pi/2$ pulse and it does not lay on the equator at the end of the rotation, yielding a distorted parity approximation that degrades dramatically with increasing $n$, as illustrated in Fig.~\ref{fig: Fig2}b. 

A standard technique to mitigate parity measurement errors from the skewed qubit rotation is to calibrate for a shorter waiting time  $t_w < \pi/\chi$ that maximizes the contrast. Another method to mitigate parity mapping errors is to perform two parity measurements in succession for each point in phase space. In this scheme, the second $\pi/2$-rotation in the sequence is enacted with opposite phases such that mapping between parity and qubit state is flipped. Finally, the corrected parity is computed by taking the difference between these two different mappings to remove the spurious phase accumulation. However, this correction demands twice the number of measurements and still results in a scaling error of the resulting parity,
\begin{equation}
    P_{\text{cor}}=(P-P_{\text{inv}})/2=\eta P_{\text{id}},
\label{eqn:imperfect_Pcor_coherent}
\end{equation}
where $P$ and $P_\text{inv}$ are the standard and inverted parity, $P_\text{id} = \text{Tr} \left( \rho e^{i \pi \hat n} \right)$ is the ideal parity, and $\eta$ is the scaling factor (Appendix~\ref{apx_sub:parity-mapping}).

Experimentally, this degradation of parity mapping with increasing excitation number in the CV state can be readily observed. We measure the parity of a series of Fock states with the standard procedure described above, using 16-ns $\pi/2$ pulses and 284-ns waiting time. The results, shown in Fig.~\ref{fig: Fig2}a, show a larger discrepancy between the experimentally measured $|P_{corr}|$ and its ideal value of 1 as the Fock state number of the cavity $|n\rangle$ increases. This distorted mapping makes parity a nonideal observable for state reconstruction, highlighting the importance of exploring the excitation number as a more reliable observable.

% While naively it seems that this distortion could be removed by using infinitely short $\pi/2$-pulses, this is unrealizable due to the finite anharmonicity of the qubit even in state-of-the-art cQED hardware.

Apart from its resilience against coherent errors stemming from the Hamiltonian itself, ORENS is also robust against qubit decoherence during the process of mapping the excitation number onto the qubit state. In general, qubit decoherence is described by the two independent mechanisms of energy decay and dephasing, which are characterized by the coherence times $T_1$ and $T_{\phi}$, respectively. While standard cQED setups can reliably achieve a $T_1$ in the range of several tens to hundreds of microseconds~\cite{kjaergaard_superconducting_2020}, ensuring a consistent $T_{\phi}$ proves to be a challenging task~\cite{gargiulo_fast_2021}. This challenge is particularly pronounced in the case of flux-tunable qubits, where $T_{\phi}$ can be as short as a few microseconds~\cite{hutchings_tunable_2017}. Considering a dephasing rate of the qubit $\Gamma_{\phi} = 1/T_{\phi}$, by solving the Lindblad master equation, the excitation number decays as
\begin{equation}
    p^{\prime}_n\equiv p^{\prime}_e(\Delta=\chi n) \approx \rho_{nn} \times \frac{1}{2}(1+e^{-\frac{t_{\pi}}{2T_{\phi}}})
    \label{eqn:pn_t2_decay}
\end{equation} (Appendix~\ref{apx_sub:qubit-decoherence}), where we have used a prime notation to denote the case under decoherence. Equation~(4) shows that $p_n^{\prime}$ contains a constant term (half of the magnitude of the ideal observable, $\rho_{nn}=\langle n|\rho|n\rangle$) and another that decays exponentially with the dephasing rate. This means that a finite proportion of the constrast is always preserved in measurement.

% This indicates that only halfof the magnitude of the ideal observable ($\rho_{nn}=\langle n|\rho|n\rangle$) decays exponentially with the dephasing rate. 

\begin{figure}[t!]
\centering
\includegraphics{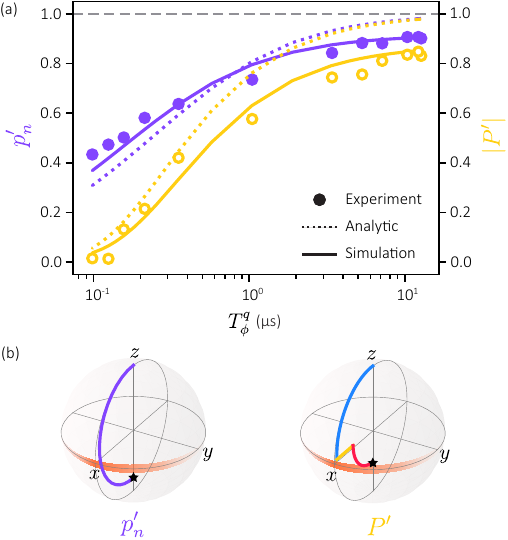} 
\caption{{\bf Excitation number and parity of the vacuum cavity state under qubit dephasing}. 
(a) The measurement outcomes of excitation number $p_n^{\prime}$ (purple) and absolute value of corrected parity $|P^{\prime}|$ (yellow) mappings for the vacuum state are plotted against varying $T_\phi$ of the qubit. They show good agreement with analytical (Eq.~(\ref{eqn:pn_t2_decay}) and Eq.~(\ref{eqn:parity_t2_decay})) and simulated trends based on real device parameters. With roughly half the contrast preserved, excitation number measurements are robust against qubit dephasing in cQED, whereas parity mapping suffers significantly more degradation. (b) Intuitive illustrations for the resilience shown by excitation number mapping. The qubit state spends less time near the equator (orange) in mapping $p_n^{\prime}$ than $P^{\prime}$, and thus is less susceptible to dephasing.
}
\label{fig: Fig3}
\end{figure}

We perform an experiment to measure $p_0$ of the vacuum cavity state at various engineered qubit dephasing time $T_{\phi}$ to observe its impact on the observable mapping (Fig.~\ref{fig: Fig3}). The $T_{\phi}$ of the qubit is shortened on-demand by using excitation-induced dephasing~\cite{sears_photon_2012}, where a weak coherent tone continuously drives the readout resonator (Appendix~\ref{apx_sub:engineering-dephasing}). To effectively isolate the error due to dephasing, the dephasing tone is only applied during the mapping of the observable, after the state preparation pulses, and with a calibrated buffer time to allow the resonator to depopulate before the readout tone. The duration $t_\pi$ of the $\pi$-pulse and the dispersive coupling $\chi$ are both fixed.

Our results show that excitation number mapping is partially preserved under qubit dephasing. At $T_\phi = 4\,\mu\text{s} = 4 t_{\pi}$, we observe a $p_{0}>0.8$, and for as low as $T_\phi =0.1\,\mu\text{s} = t_{\pi}/10$, the expectation value remains above 0.4. To benchmark this result, we repeat the same protocol with parity mapping. From the Lindblad master equation, we find that the whole parity observable exponentially decays to 0 (i.e. total loss of information) as $T_{\phi}$ goes to 0. Analytically, this is described by
\begin{equation}
     P^{\prime}\equiv 2p_e^{\prime}-1 \approx P_{\text{id}}\times e^{-\frac{t_{w}}{T_{\phi}}},
    \label{eqn:parity_t2_decay}
\end{equation}
where $t_w$ denotes the variable waiting time between the two $\pi/2$ pulses (Appendix~\ref{apx_sub:qubit-decoherence}). Since parity mapping relies on the qubit state acquiring a deterministic phase, low dephasing times $T_\phi < 0.1\,\mu$s completely erase this mapping, as shown in Fig.~\ref{fig: Fig3}a. At high $T_{\phi}$, we expect both excitation number and parity of the vacuum state to saturate the ideal value in the absence of coherent errors. 
However, errors due to parasitic dynamics become dominant in the parity-mapping process for states with more excitations, as shown in Fig.~\ref{fig: Fig2}. This leads to more pronounced reconstruction imperfections. Thus, it is apparent that excitation number mapping offers a more robust performance than the standard parity measurements against both coherent errors and decoherence.

% \textcolor{blue}{We also note that in Fig.~\ref{fig: Fig3}, where measurements were conducted on vacuum cavity state, indeed with improvements in $T_{\phi}$ we expect that both photon number and parity would saturate the ideal value. However, this is valid only for vacuum cavity state where there is no coherent error. For states with higher excitations, coherent errors will become dominant (shown in Fig.~\ref{fig: Fig2}) regardless of $T_{\phi}$. Therefore, if $T_{\phi}$ is high, we expect that the photon number $p_n^{\prime}$ is even closer to the ideal values, while parity suffers from coherent error.} 

% We have thus demonstrated that excitation number mapping offers a more robust performance than the standard parity measurements, indicating it is the more favorable observable to measure compared to existing state reconstruction techniques. With this, we can now use the ORENS technique to verify its ability to efficiently and accurately reconstruct density matrices.

\section{ORENS STATE RECONSTRUCTION}

To fully demonstrate the efficacy of ORENS, we use it to perform full CV state reconstruction for states of different sizes in our bosonic cQED system. For a given Hilbert space of dimension $D$, we first obtain the set of optimized displacements by sweeping over the excitation number $n \in [1, D-1]$ and implementing a gradient-descent algorithm to find $\{\alpha_k\}_{k=1}^{N^*_{obs}}$, where the cost function to minimize is the condition number of the measurement matrix (Appendices~\ref{apx_sub:linear-inversion} and \ref{apx_sub:optimizing-set-measurements}). This set of measurements allows us to reconstruct any arbitrary state bounded by dimension $D$. To benchmark the performance of ORENS comprehensively, we prepare all Fock states $|k\rangle$ and their superpositions $|j\rangle + e^{i \phi}|k\rangle/\sqrt{2}$ with $j<k = 0, \cdots, D-1$ and $ \phi = \{0, \pi/2\}$ ($D^2$ different states in total) within the same truncation dimension. Experimentally, each state is prepared by playing $2$-$\mu$s numerical pulses after a qubit pre-selection measurement, which eliminates the residual thermal population ($\approx 2\%$). Next, we displace the cavity state in phase space with the set of optimized displacements, $\hat D(\alpha_k)$, via a drive resonant at the cavity frequency with a Gaussian envelope. As a concrete example, the optimal displacements for $D=6$ are 35 unique $\{\alpha_k \}_{k=1}^{35}$ each followed by a measurement of excitation number $n=5$ in this example. This is enacted by exciting the qubit with a pulse resonant at the qubit frequency conditioned on the cavity having $n$ excitations, $\omega_q - n \chi$, where $n=5$. The resulting qubit state is obtained through standard single-shot dispersive readout. These measurement outcomes are processed to estimate the cavity state, first by inverting Born's rule to obtain the least-squares estimator $\rho_\text{LS}$, and then by using Bayesian inference~\cite{lukens_practical_2020} to obtain $\rho_\text{BME}$ as the final reconstructed estimator. The Bayesian method treats uncertainty in meaningful ways and utilizes all available information optimally. As a result, this approach affords the most faithful estimator for the state~\cite{blume-kohout_accurate_2006, blume-kohout_optimal_2010}, particularly in comparison to the typical maximum likelihood estimation approach~\cite{smolin_efficient_2012}. More details on the Bayesian interference technique used in this study are presented in Appendix~\ref{apx_sub:bayesian}.

%this set of displacements for a given truncation dimension $D$, we sweep over the excitation number $n \in [1, D-1]$, where for each set we run a gradient-descent algorithm over the set of displacements $\{\alpha_k\}_{k=1}^{N^*_{obs}}$ to minimize the condition number of the measurement matrix (Appendix~\ref{apx_sub:linear-inversion} and Appendix~\ref{apx_sub:optimizing-set-measurements}).

\begin{figure*}[t!]
\centering
\includegraphics[width=1.0\textwidth]{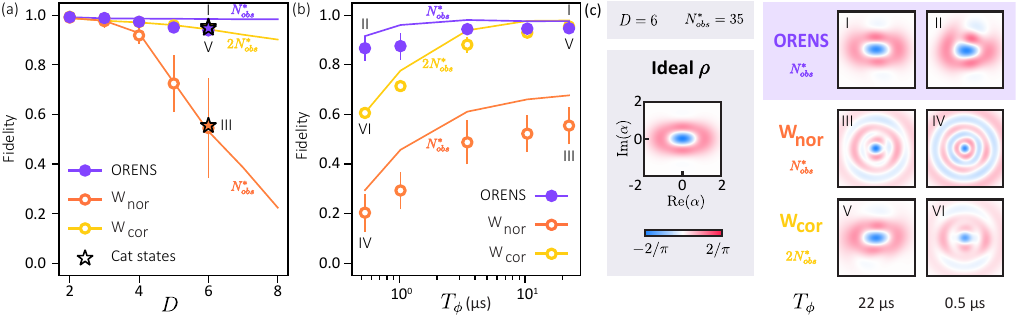}
\caption{{\bf State reconstruction for ORENS ($N^*_{obs}$), normal Wigner ($N^*_{obs}$), and corrected Wigner ($2N^*_{obs}$).}
(a) The average reconstruction fidelities for $D^2$ different states (Fock states and their superpositions) against cut-off dimension $D$. They show good agreement with simulated trends (solid curves) based on real device parameters.
The star markers are average fidelities for reconstructing four small cat states, well contained within $D=6$. Beyond $D=6$, there are no meaningful experimental points as the readout was distorted by cross-Kerr between the readout resonator and cavity. The qubit dephasing time here is $T_{\phi}=15.3~\mu$s. (b) The average reconstruction fidelities of the four small cat states for different qubit dephasing times $T_{\phi}$. (c) Wigner functions of the reconstructed odd cat states ($|\alpha\rangle - |-\alpha\rangle$) with ORENS, normal Wigner, and corrected Wigner techniques at two different $T_{\phi}$ points. With strong retention of important cat state properties, ORENS outperforms both Wigner reconstruction techniques with the fewest measurements, particularly at high dimensions and under qubit dephasing.
}
\label{fig: Fig4}
\end{figure*}

To evaluate the quality of the state reconstruction, we compute the fidelity between the estimated density matrix and the target density matrix $\rho_{\text{tar}}$, generated by simulating the GRAPE pulses under decoherence,
\begin{equation}
    F = \left( \text{Tr}\sqrt{\sqrt{\rho_{\text{tar}}}\rho_\text{BME} \sqrt{\rho_{\text{tar}}}}\right)^2.
\end{equation}
The average fidelity over the $D^2$ different states for each dimension is plotted in Fig.~\ref{fig: Fig4}a, up to $D=6$. Across all dimensions, the reconstruction fidelity using ORENS exceeds 95\%. Beyond $D=6$, the readout was distorted due to the dominant cross-interaction between the cavity and the readout resonator present for larger bosonic states, resulting in the absence of meaningful experimental points. This is not a fundamental limitation of the technique but a rather device-specific artifact. The dominant error mechanisms impacting the reconstruction fidelity are the thermal population of the qubit and imperfect single-shot readout discrimination (Appendix~\ref{apx_sub:thermal-population},~\ref{apx_sub:error-budget}), which could be significantly reduced by improving device thermalization and using a quantum-limited amplifier, respectively.  

Having tested the reconstruction technique on Fock states and their superpositions, we evaluate the protocol for cat states, a versatile backbone of CV information processing protocols~\cite{joo_quantum_2011, ralph_quantum_2003, chamberland_building_2022, ofek_extending_2016, puri_bias-preserving_2020}. We repeat the same reconstruction protocol for $D=6$ with four small cat states: 
$|\alpha\rangle \pm |-\alpha\rangle$ and $|\alpha\rangle \pm i |-\alpha\rangle$ (normalisation implied), with $\alpha = 1$. The averaged fidelity matches the Fock state superposition fidelities, see the purple star in Fig.~\ref{fig: Fig4}.

We benchmark the reconstruction performance of both Fock and cat states against two versions of the standard Wigner protocol, normal and corrected. For both variations, we fix the measurement observable to parity and optimize the corresponding $N^*_{obs}$ displacements. For corrected Wigner, the distortion due to the always-on dispersive interaction is partially removed at the cost of increasing the number of independent measurements to $2N^*_{obs}$, which is twice the theoretical minimum. 

With a focus on using only the minimal number of measurements $N^*_{obs}$, ORENS demonstrates accurate reconstruction across dimensions, significantly surpassing the performance of the normal Wigner protocol. While the corrected Wigner tomography affords comparable state reconstruction fidelity to ORENS, but it is more resource-intensive requiring double the number of independent measurements. Compared to prior works using generalized Q function measured with a grid of unoptimized displacements and all the Fock state projections up to the maximum photon number~\cite{kirchmair_observation_2013}, ORENS affords a $50$-fold reduction in the number of measurements. 

To further verify the performance of ORENS under qubit dephasing, we prepare the same four cat states and reconstruct them using ORENS for the several independently calibrated qubit $T_{\phi}$ points as shown in Fig.~\ref{fig: Fig4}b. Remarkably, we notice only a slight deterioration of the average reconstruction fidelity, with fidelities exceeding $>86\%$. This demonstrates that even when using a conditional $\pi$-pulse duration that far exceeds the qubit $T_{\phi}$, the excitation number observable still maps sufficient information to accurately reconstruct states. 

The robustness under dephasing attests to the versatility of the ORENS in regimes of low-$\chi$ between the cavity and the qubit. To maintain an equivalent frequency-selectivity, a smaller $\chi$ demands a longer $t_\pi$ (Appendix~\ref{apx_sub:ens-mapping-selectivity}). Considering our experimental (simulated) fidelity of 88\% (92\%) with $\chi/2\pi = 1.4$~MHz and $T_{\pi}/ T_{\phi} = 1\mu\text{s}/0.5\mu\text{s} = 2$, we can expect an equivalent fidelity with $\chi/2\pi = 35$~kHz, $T_{\pi}/ T_{\phi} = 40\mu\text{s}/20\mu\text{s} = 2$. This was verified with a simulation to reconstruct the cat states, with an average fidelity of 94\%.

\section{CONCLUSION AND DISCUSSION}
Through the above analytic and experimental results, we have demonstrated:
(1) a powerful technique for Optimized Reconstruction with Excitation Number Sampling (ORENS) with minimal measurements that relies only on displacements and excitation counting, and can be readily applied across CV experimental platforms, 
(2) clear evidence that excitation number mapping in bosonic cQED is an ideal and convenient observable for state reconstruction that can be directly implemented on standard devices without any tailored operations or parameters,
(3) the robustness of excitation number mapping even under severe qubit dephasing in cQED, and 
(4) the ability of ORENS to reliably reconstruct arbitrary states of all dimensions in the presence of pronounced coherent and incoherent errors .

% \section{\label{sec:level1}SUMMARY AND DISCUSSION}
For each of the experiments, ORENS outperforms the state-of-the-art Wigner reconstruction with the fewest measurements. Although the fidelities obtained with ORENS are nearly matched by the corrected Wigner strategy, our method uses half the number of measurements and scales more favorably with state dimensionality. The primary drawback of ORENS in cQED is the high sensitivity to undesired residual excitations of the qubit (Appendix~\ref{apx_sub:thermal-population}). However, the reconstruction fidelity can be readily improved with good thermalization of the qubit as well as standard pre-selection measurements. 

Looking beyond, the ORENS technique can be readily implemented for multimode systems. For example, for a two-mode system with modes $A$ and $B$, each with dimension $D$, we would apply the displacements $\{D_A(\alpha_{i})\otimes \openone_B, \openone_A\otimes D_B(\alpha_{j}), D_A(\alpha_{i})\otimes D_B(\alpha_{j})\}$, where $\{\alpha_{i}\}$ is the set of optimized single-mode displacement points, before measuring the joint excitation numbers. In cQED, the joint excitation numbers can be directly extracted by either using a single ancillary qubit coupled to both cavities via a selective $\pi$ pulse tuned to $\omega_q - n_A\chi_A - n_B\chi_B$, or by using two single-cavity ancillary qubits and correlating their outcomes.
Not only would this multimode reconstruction approach likely outperform the standard joint Wigner tomography for similar arguments of excitation number observable robustness, but the measurement of joint excitation number is significantly more convenient than joint parity. Joint parity measurements are challenging, as they require either designing $\chi_A=\chi_B$ or utilizing higher levels of the transmon with concatenated single-mode conditional phase gates~\cite{wang_schrodinger_2016}. While the generalized Wigner function~\cite{chakram_multimode_2022} helps overcome these practical challenges, the arbitrary relative phase of the modified Ramsey sequence tends to reduce the contrast of measurement outcomes and renders the reconstruction less robust.

Overall, we have developed and demonstrated a versatile technique to efficiently and robustly estimate arbitrary CV states, accessible across different bosonic hardware platforms. Our results bring us one step closer to scalable and reliable characterization and verification across CV quantum applications.\\ 

\FloatBarrier

\begin{acknowledgments}
WE acknowledge the funding
support of the National Research Foundation grant number NRF2020-NRF-ISF004-3540 and the Ministry of Education, Singapore grant number MOE-T2EP50121-0020.
T.K. thanks Tomasz Paterek for discussions during the earlier stages of this project.
\end{acknowledgments}

\section*{Data and Codes availability}
\label{apx:github-code}
All data and codes needed to evaluate the conclusions of the paper are available on GitHub:
\url{https://github.com/clarayfontaine/ORENS_bosonic_state_reconstruction}.

%\clearpage
\appendix
\section{Experimental Device}
\label{apx:experimental-device}
\subsection{Design and tools}
\label{apx_sub:design-tools}
The experimental device used in this work is a standard bosonic cQED system~\cite{blais_cavity_2004, girvin_circuit_2014} in the strong dispersive-coupling regime. It consists of a superconducting microwave cavity, dispersively coupled to a transmon qubit for controllability and readout, which is also dispersively coupled to a planar readout resonator. 
We simulate the electromagnetic fields of the device using Ansys finite-element High-Frequency Simulation Software (HFSS) and obtain the Hamiltonian parameters using the energy participation ratio (EPR) approach~\cite{minev_energy-participation_2021}.
The key system properties, such as the frequency of each circuit and the pair-wise non-linear couplings between them, are iteratively refined to meet the target parameters.
In this section, we describe the details of the design considerations, the resulting properties of the main elements in the device, and the main calibration procedures for the different experimental parameters.

\subsection{Package and chip fabrication}
\label{apx_sub:package_chip_fab}
The storage cavity is a three-dimensional high-Q coaxial $\lambda/4$-resonator with a cut-off frequency $f_\text{cut} \sim 600$ MHz.
The cavity and the coaxial waveguide that hosts the qubit and the resonator are machined out of high-purity (4N) aluminum, where the external layer (~0.15 mm) has been removed with chemical etching to reduce fabrication imperfections.
The ancillary transmon qubit and the planar readout resonator are fabricated by evaporating aluminum on a sapphire substrate.
The design is patterned using a Raith electron-beam lithography machine, on a HEMEX sapphire substrate (i.e. sapphire substrate grown by the heat exchanger method graded based on superior optical properties) cleaned with 2:1 piranha solution for 20 minutes and coated with 800~nm of MMA and 250~nm of PMMA resist. The pattern is then developed with a mixture of de-ionized water and isopropanol at a 3:1 ratio. Using a PLASSYS double-angle evaporator we deposit the two aluminum layers of 20 nm and 30 nm thickness at -25 and +25 degrees, respectively, separated by an oxidation step with a mixture of 85\% $O_2$ and 15\% Argon at 10 mBar for 10 minutes. The chip is finally diced on an Accretech machine and inserted in the waveguide, with an aluminum clamp where we use indium wire to improve thermalization.

\subsection{Intrinsic Purcell filtering}
\label{apx_sub:purcell}
The design of the Hamiltonian parameters considers the ORENS protocol requirements and the versatility to explore different decoherence regimes.
We design the dispersive interaction between the cavity and the qubit to be $\sim 1.4$ MHz. Hence, our selective pi-pulses need to be $\sim 1 \mu$s long, and the Ramsey revival time $\sim \pi/\chi \approx 3.14$~$\mu$s.

\begin{figure}[h]
    \centering
    \includegraphics{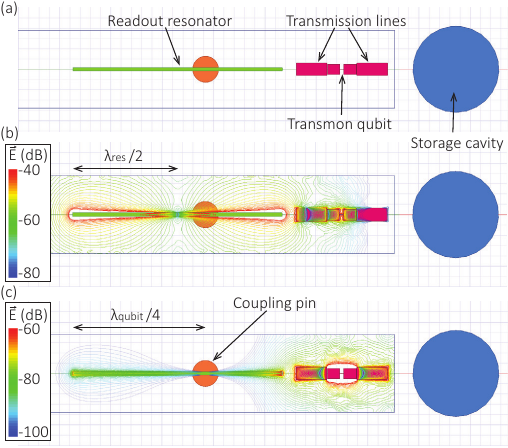}
    \caption{\textbf{Chip design.} a) Simulated readout resonator, qubit and cavity modes in Ansys HFSS. b) and c) show the complex magnitude of the electric field of the resonator and qubit modes, respectively. The position of the coupling pin (orange cicrcle) is at a distance $\lambda_\text{qubit}/4$ from the end of the resonator, where the qubit field is minimum, while the resonator field is non-zero.}
    \label{fig:chip}
\end{figure}

To achieve long-enough coherence times for the qubit, we mitigate its resonator-mediated Purcell decay by designing an intrinsic Purcell-filter structure~\cite{sunada_fast_2022}, see Fig.~\ref{fig:chip}. This is done by optimizing the position of the coupled transmission line. Simulations show that the optimal position aligns with the voltage node of the qubit field at approximately $\lambda_\text{qubit}/4$ away from the end of the resonator. In this position, the qubit field is very weakly coupled to the transmission line while the readout resonator is significantly coupled for fast readout. 

For ease of fabrication and to preserve cavity coherence, the transmission line is kept at an appropriate distance. To satisfy this requirement, as well as the strong dispersive coupling and the intrinsic-purcell filtering condition at the same time, we added two planar stripline-like structures on both ends of the qubit pads. The strips are short enough not to introduce any mode below 8 GHz, while effectively guiding the transmon field to the cavity and to readout resonator modes as desired.

\subsection{Hamiltonian parameters and coherence times}
\label{apx_sub:hamiltonian}
Expanding the cosine term of the Josephson junction up to the fourth order, we can write the full Hamiltonian of our system as
\begin{equation}
\begin{split}
    \frac{\hat{{H}}}{\hbar} &= \omega_\text{c} \hat{{c}}^\dagger \hat{{c}}
        +\omega_\text{q} \hat{{q}}^\dagger \hat{{q}} 
        +\omega_\text{r} \hat{{r}}^\dagger \hat{{r}} \\
        &- \frac{\chi_\text{cc}}{2} \hat{{c}}^\dagger \hat{{c}}^\dagger \hat{{c}} \hat{{c}}
        - \frac{\chi_\text{qq}}{2} \hat{{q}}^\dagger \hat{{q}}^\dagger \hat{{q}} \hat{{q}}
        - \frac{\chi_\text{rr}}{2} \hat{{r}}^\dagger \hat{{r}}^\dagger \hat{{r}} \hat{{r}} \\
        &- \chi_\text{cq} \hat{{c}}^\dagger \hat{{c}} \hat{{q}}^\dagger \hat{{q}}
        - \chi_\text{qr} \hat{{q}}^\dagger \hat{{q}} \hat{{r}}^\dagger \hat{{r}}
        - \chi_\text{cr} \hat{{c}}^\dagger \hat{{c}} \hat{{r}}^\dagger \hat{{r}},
\end{split}
\label{eq: Ham}
\end{equation}
where $\omega_{\text{i}}$ and $\hat \imath$ respectively denote the angular frequency and annihilation operator of the system with $i={c}, {q}$, and ${r}$ corresponding to cavity, qubit, and resonator.
% where the $\omega_c, \omega_q, \omega_r$ respectively denotes the angular frequencies of the cavity, qubit, and resonator, and 
The $\chi_{ij}$ of the second and third lines correspond to the self-Kerr and cross-Kerr interactions between modes, respectively. 
The value of the experimentally calibrated parameters can be seen in tables~\ref{tab:frequencies} and~\ref{tab:kerrs}.
The high self-Kerr of the transmon allows effective treatment of it as a qubit with two energy levels $|g\rangle$ and $|e\rangle$, where $\hat q$ ($\hat q^{\dagger}$) can be replaced with $|g\rangle \langle e|$ ($|e\rangle \langle g|$).

\begin{table}[h]
    \centering
    \begin{tabular}{ P{1.5cm}|P{1.7cm}|P{1.5cm}|P{1.5cm}|P{1.5cm} }
    % \hline
    & $\omega / 2\pi $(GHz) & $T_1 (\mu$s) & $T_2 (\mu$s) & $T_2^\text{echo} (\mu$s)\\
    \hline
    Qubit   & 5.277 & 85-113  & 14-22  & 44-48 \\
    \hline
    Cavity  & 4.587 & 992   & - & -\\
    \hline
    Resonator & 7.617 & 2.08  & - & -\\
    % \hline
    \end{tabular}
    \caption{Frequency and coherence times for the 3 modes of the device.}    
    \label{tab:frequencies}
\end{table}

\begin{table}[h]
    \centering
    \begin{tabular}{P{1.5cm}|P{2cm}|P{2cm}|P{2cm}}
     % \hline
      & Cavity & Qubit & Resonator\\
     \hline
     Cavity & 4-6 kHz & 1.423 MHz & 2 kHz \\
     \hline
     Qubit & 1.423 MHz & 175.3 MHz  &  0.64 MHz \\
     \hline
     Resonator & 2 kHz &  0.64 MHz & - \\
     % \hline
    \end{tabular}
    \caption{\textbf{Table of Kerr interactions.} Diagonal elements correspond to the self-Kerr interactions of each mode, and off-diagonal terms correspond to the cross-Kerr interactions between different modes.}
    \label{tab:kerrs}
\end{table}

The next major contribution to Eq.~(\ref{eq: Ham}), stemming from the sixth-order term of the cosine expansion, corresponds to the second-order dispersive interaction between the cavity and the qubit, 
$ -\chi'_\text{cq} q^\dagger q c^\dagger c^\dagger c c $. 
By fitting the resonance frequencies of the qubit to second order on the number of excitations in the cavity, we find 
$ \chi'_\text{cq} /2\pi \approx 16$ kHz.

\subsection{Microwave wiring}
\label{apx_sub:microwave-wiring}
\begin{figure}[h!]
    \centering
    \includegraphics{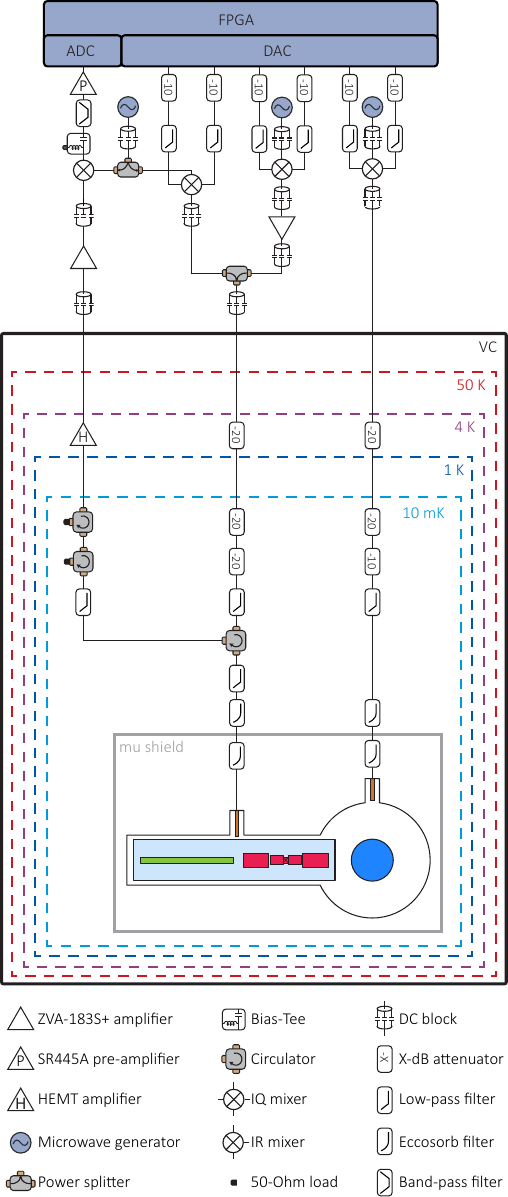}
    \caption{\textbf{Experimental setup.} Schematic of the RF components and connections at room temperature and inside the Bluefors dilution refrigerator.}
    \label{fig:fridge-wiring}
\end{figure}

The radio-frequency (RF) pulses to drive the readout resonator, qubit, and cavity are created by IQ-mixing the local oscillator (LO) signal from a Vaunix Lab Brick microwave resonator with the intermediate-frequency (IF) $\mathcal{I}$ and $\mathcal{Q}$ quadratures generated by the Digital-to-Analogue Converter (DAC) port of a Quantum Machines fast field-programmable gate array (FPGA). 
For the readout, we measure the reflected signal from the resonator, which is amplified in a High-electron mobility transistor (HEMT) amplifier and a room-temperature ZVA-183S+ amplifier before being down-converted to 50~MHz with a Marki IR-mixer. The signal is finally amplified with a Stanford Research Systems SR445A room-temperature amplifier before being sampled in the Analogue-to-Digital Converter (ADC) block of the FPGA. The schematic wiring setup can be in seen Fig.~\ref{fig:fridge-wiring}.

\begin{figure*}[!]
    \centering
    \includegraphics{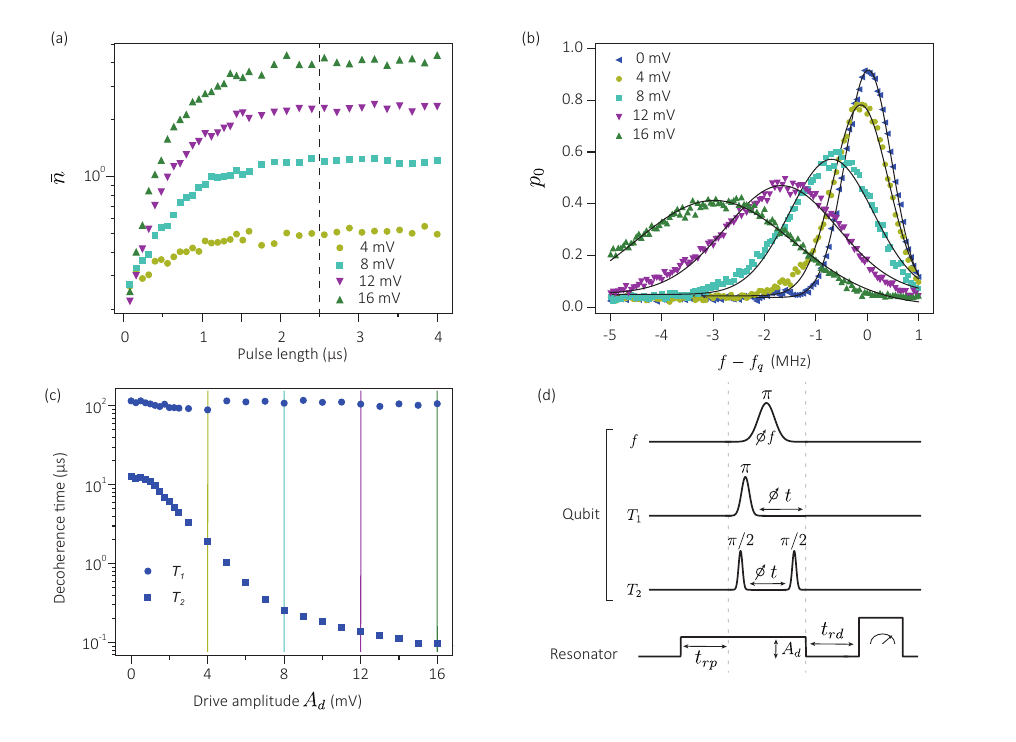}
    \caption{\textbf{Engineering qubit dephasing.} (a) Average excitation number $\bar{n}$ in the resonator as a function of the various pulse length for several drive amplitudes $A_d$. The resonator population reaches a steady state after a ring-up time $t_{rp}$ for all $A_d$, which is taken as 2.5~$\mu$s (black dotted line). (b) Qubit spectra after driving the resonator to steady state with square pulses of varying drive amplitudes. Black lines represent Gaussian fits. The measurement pulse sequence is shown in (d), where the ring-up time $t_{rp}$ and the ring-down time $t_{rd}$ are both 2.5~$\mu$s, and the length of the $\pi$ pulse is 1~$\mu$s. (c) Qubit energy relaxation time $T_{1}$ and dephasing time $T_{2}$ obtained with different the resonator drive amplitude. The vertical lines correspond to the drive amplitudes used in (a) and (b). The qubit $T_1$ and $T_2$ times in the presence of photons in the readout resonator are obtained with the pulse sequences in (d).
    } 
    \label{fig:badt2_details}
\end{figure*}

\subsection{Engineering qubit dephasing}
\label{apx_sub:engineering-dephasing}
To demonstrate the robustness of ORENS under qubit dephasing, we engineer the dephasing time $T_\phi$ of the qubit. This is achieved by driving the dispersively-coupled readout resonator to a steady-state photon population that induces dephasing via photon-shot noise. The dephasing rate is controlled by varying the average excitation number in the resonator~\cite{sears_photon_2012}. 

% we can control the rate at which the qubit experiences dephasing
% events by varying in situ the cavity mode population and decay rate
% The photon shot noise in the resonator induces qubit dephasing in the dispersive regime, and the dephasing rate of qubit exhibits a linear relationship with the average excitation number within the resonator. In experiment, we control the dephasing rate of the qubit by driving the resonator to steady state with a corresponding calibrated average excitation number.
    
To calibrate the average excitation number $\bar{n}$ in the resonator as a function of the drive amplitude, we populate the resonator with a square pulse and conduct a Ramsey experiment. Subsequently, we fit the modulated Ramsey oscillations using the free parameter $\bar{n}$ ~\cite{mcclure_rapid_2016}. For a given drive amplitude $A_d$ -- expressed as the voltage of the DAC output --, we can extract $\bar{n}$ as a function of the pulse length, see Fig.~\ref{fig:badt2_details}a.
For all drive amplitudes, the resonator reaches a steady state after 2.5~$\mu$s, which is set as the ring-up time $\tau_{rp} = 2.5~\mu$s. Following a similar calibration, we choose the ring-down time $\tau_{rd}$ for resonator decay as 2.5~$\mu$s.

% The maximal drive amplitude $A_d$ in Fig. 4(b) in the maintext is 6 mV, so the corresponding maximum excitation number in the resonator is lower than 1. Consider such a small excitation number and the coupling between the cavity and the resonator is 2kHz, so the cross-Kerr effect is minor.

Driving the resonator not only induces qubit dephasing but also shifts the qubit frequency due to its dispersive interaction. As shown in the qubit spectroscopy plot in Fig.~\ref{fig:badt2_details}b measured with the sequence in d, the qubit peak broadens and shifts with increasing drive amplitude $A_d$. To isolate the dephasing effect from the frequency-shifting effect in the experiments that follow, we perform the observable mapping protocols using the shifted frequency corresponding to the resonator drive amplitude. 

To characterize the qubit's pure dephasing time $T_\phi = 1 / (1 / T_2 - 0.5 / T_1)$ for each resonator drive, we measure the qubit energy relaxation time $T_1$ and qubit dephasing time $T_2$ after driving the resonator to steady state and updating the qubit frequency (pulse sequences in Fig.~\ref{fig:badt2_details}d). As shown in Fig.~\ref{fig:badt2_details}c, $T_1$ stays relatively constant with increasing resonator drive amplitude, while $T_2$ decreases smoothly. This indicates the photon-shot noise in the resonator only induces qubit dephasing, not qubit energy relaxation.

\begin{figure}[!]
    \centering
    \includegraphics{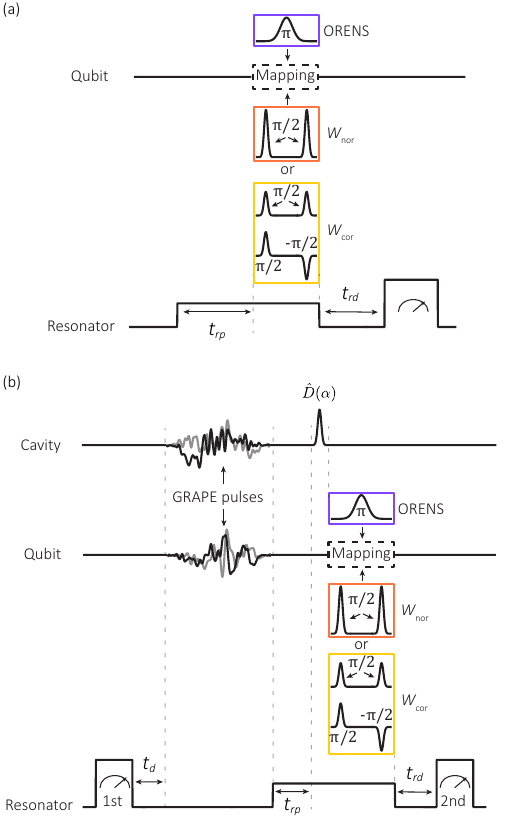}
    \caption{\textbf{Observable mapping sequences with engineered qubit dephasing.} (a) Pulse sequence for ORENS and Wigner observable mapping (Fig.~\ref{fig: Fig3}). The ring-up time $t_{rp}$ and the ring-down time $t_{rd}$ are both 2.5~$\mu$s, the lengths of the $\pi$ pulse and the $\pi/2$ pulse are 1~$\mu$s and 64~ns respectively, and the delay time $\tau$ for parity measurement is 284~ns. (b) The pulse sequence ORENS and Wigner state reconstruction (Fig.~\ref{fig: Fig4}(b) and (c)). We apply an additional measurement pulse to the resonator at the start for post-processing selection. The delay time $t_d$ for the resonator decay after the first measurement, the ring-tp time $t_{rp}$, and the ring-down time $t_{rd}$ are both 2.5~$\mu$s. The length of the GRAPE pulses is 2~$\mu$s, the length of the displacement pulse $D(\alpha)$ is 240~ns, the lengths of the $\pi$ pulse and the $\pi/2$ pulse are 1~$\mu$s and 64~ns respectively, and the delay time $\tau$ for parity measurement is 284~ns.}
    \label{fig:pulse sequences}
\end{figure}

Having calibrated how to engineer $T_\phi$, we now study how the measurement observables -- excitation number $p_n$ for ORENS and parity $P$ for Wigner -- behave when increasing the qubit dephasing rate while keeping the cavity in vacuum state, see Fig.~\ref{fig: Fig3} in the main text. The pulse sequence is shown in Fig.~\ref{fig:pulse sequences}a. For a certain calibrated $T_\phi$, we drive the resonator to steady state, update the qubit frequency, and conduct the standard $p_n$ and $P$ observable mapping protocols. Then, after a ring-down time $t_{rd}$ for the resonator to de-populate, the measurement pulse is applied. 

We evaluate ORENS and Wigner state reconstruction under dephasing, see Fig.~\ref{fig: Fig4} in the main text, using the pulse sequence in Fig.~\ref{fig:pulse sequences}b. To mitigate the error due to the qubit thermal population, we first measure the qubit state to later post-select the data. After waiting a delay time $t_d$ for the resonator to decay, we prepare the small cat states using GRAPE pulses with a length of 2~$\mu$s. After state preparation, we drive the resonator to steady state to induce a reduced $T_\phi$, apply a displacement pulse to the cavity, and perform either the $p_n$ or $P$ mapping. Then, we turn off the resonator drive and wait a ring-down time $t_{rd}$ for the resonator decay, before applying a final measurement pulse.

\subsection{Error budgeting}
\label{apx_sub:error-budget}
% I use data from PNS experiments on Nov 3
% Thermal pop 1.9%
% P(g|e) = 3.7%
% P(e|g) = 3.3%
We use a standard square low-power readout pulse at the resonance frequency of the readout resonator when the qubit is in the ground state. The length of the readout pulse is 1.5 $\mu$s and the reflected signal is acquired for a total time of $2.4$~$\mu$s. We measure a readout fidelity $F_\text{RO} = 1 - (p(e|g)+p(g|e))/2 = 96.5\%$, of which we estimate an infidelity of 1.9\% due to thermal population, 1.4\% due to readout discrimination error (overlap), and 0.4\% due to qubit decay during the readout. 

\begin{figure}[h!]
    \centering
    \includegraphics[scale=1]{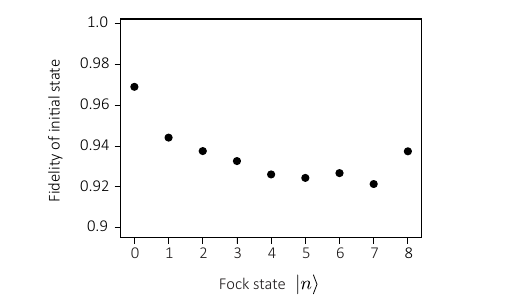}
    \caption{\textbf{Quality of state preparation.} Fidelities of the Fock states created using numerically optimized pulses.}
    \label{fig: GRAPE_fids}
\end{figure}

The cavity was measured to have a 3\% residual population, i.e. $p_1=3\%$.
% Pe after preselection = 0.9 \% 
The cavity states were prepared with numerical pulses optimized with the GRAPE algorithm applied simultaneously at the qubit and the cavity, with a fixed total duration of $2$~$\mu$s, for all states.
The average residual thermal population of the qubit after the GRAPE pulses is measured to be 4.9\%, and is further attenuated down to 3\% by performing a 1.5$~\mu$s-long readout before the numerical pulses and pre-selecting only those runs where the qubit was measured in the ground state. 
We calibrated a 2.5$~\mu$s-long buffer time 
% 900ns from RO pad + 400cc waiting time
between the readout pulse and the GRAPE pulses for the resonator to de-populate.
We test the quality of the state preparation by simulating the effect of the numerical pulses with the whole qubit-cavity Hamiltonian accounting for decoherence and thermal populations. The fidelity in relation to the ideal pulses can be seen in Fig.~\ref{fig: GRAPE_fids}. The 3\% infidelity for preparing the cavity in vacuum (Fock 0 in the figure) can be explained by the previously-mentioned thermal population of the cavity, and it matches the discrepancy between the experimental data and the ideal value in Fig.~\ref{fig: Fig2} in the main text. In addition, the experimental $p_n$ data shows a slow systematic decay as a function of Fock state, which can be explained by state preparation errors of the GRAPE pulses, see 
Fig.~\ref{fig: GRAPE_fids}. 

The additional loss of contrast in Fig.~\ref{fig: Fig3} compared to Fig.~\ref{fig: Fig2} is caused by qubit decoherence during the additional ring-up and ring-down times, as explained in Section~\ref{apx_sub:engineering-dephasing}.

%%%%%%%%%%%%%%%%%%%%%%%%%%%%%%%%%%%%%%%%%%%%%%%%%%%%
\section{Coherent error in observables}
\label{apx:coherent-error}
\subsection{Excitation number mapping}
\label{apx_sub:ens-mapping-selectivity}

Here we derive the expression for the qubit probability following the excitation number mapping protocol, show how it maps to excitation number of the cavity state, and discuss its selectivity.
We set $\hbar=1$ henceforth.

First, we note that given the initial state of the qubit $|g\rangle$ and its evolution governed by a Hamiltonian $\hat H=\Delta |e\rangle \langle e|+(\Omega/2)\hat \sigma_y$, the probability of finding the qubit in the excited state at a time $t$ is
\begin{equation}\label{EQ_Qpe}
    p_e(t) = \frac{\Omega^2}{\Omega^2+\Delta^2}\sin^2(\sqrt{\Omega^2+\Delta^2}\:\frac{t}{2}).
\end{equation}
At a time $t_{\pi}\equiv \pi/\Omega$, the probability is maximum ($p_e=1$) for $\Delta=0$ and decays as the detuning $|\Delta|$ increases.

To map the excitation number of the cavity onto the qubit, the sequence starts with the qubit in $|g\rangle$ and cavity in an arbitrary state $\rho$, which evolve under the Hamiltonian
\begin{equation}
    \hat H=\Delta |e\rangle \langle e|+\frac{\Omega}{2}\hat \sigma_y-\chi |e\rangle \langle e| \otimes \hat n,
\end{equation}
where $\hat n=\hat c^{\dagger}\hat c$.

The qubit state at time $t$ follows
\begin{eqnarray}\label{EQ_qstateENS}
    \rho_q(t) &=& \text{Tr}_{\text{cav}}(e^{-i\hat Ht}(|g\rangle \langle g |\otimes \rho) \:e^{i\hat Ht})\nonumber \\
    &=&\sum_n \langle n| e^{-i\hat Ht}(|g\rangle \langle g |\otimes \rho) \:e^{i\hat Ht} |n\rangle \nonumber \\
    &=&\sum_n \rho_{nn} \:e^{-i\hat H_nt}|g\rangle \langle g|  \:e^{i\hat H_nt} ,
\end{eqnarray}
where $\text{Tr}_{\text{cav}}$ is partial trace with respect to the cavity state, $\rho_{nn} \equiv \langle n|\rho|n \rangle$ denotes the diagonal elements (excitation number) of $\rho$, and $\hat H_n\equiv \Delta_n |e\rangle \langle e|+(\Omega/2)\hat \sigma_y$ now acts only on the qubit with $\Delta_n \equiv \Delta-\chi n$.

The probability of the qubit being in the excited state at a time $t_{\pi}=\pi/\Omega$ is
\begin{eqnarray}\label{EQ_ENSpe}
    p_e &=& \sum_n \rho_{nn} |\langle e|\:e^{-i\hat H_nt_{\pi}}|g\rangle|^2 ,\nonumber \\
    &=&\sum_n \rho_{nn} \: \frac{\Omega^2}{\Omega^2+\Delta_n^2} \sin^2(\sqrt{\Omega^2+\Delta_n^2}\:\frac{t_{\pi}}{2}),
\end{eqnarray}
where we have made use of Eq.~(\ref{EQ_Qpe}).
Each component in the summation of Eq.~(\ref{EQ_ENSpe}) features a function that peaks at $\Delta_n=\Delta-\chi n=0$ to a value $\rho_{nn}$. For each component, the maximum with respect to the detuning $\Delta$ is different and, if the peaks are sharp enough ($\chi/\Omega \gg 1$), different peaks do not overlap with each other, and we say that the excitation number sampling is selective.
The qubit probability can then be approximated as 
\begin{equation}
    p_e(\Delta=\chi n, t_{\pi})\approx \rho_{nn},
\end{equation}
which is the basis for the mapping of excitation number.

Eq.~(\ref{EQ_ENSpe}) can also be written as 
\begin{equation}
    p_e=\sum_n \rho_{nn} \frac{1}{1+\eta^2}\sin^2(\sqrt{1+\eta^2}\:\frac{\pi}{2}),
    \label{eq: eta}
\end{equation}
by defining the parameter $\eta \equiv \Delta_n/\Omega=(\Delta-\chi n)t_{\pi}/\pi$.
% \tk{From Eq.~\ref{eq: eta} it is clear that the contrast is solely determined by the cavity state and the parameter $\eta$. Hence, a lower $\chi$ (less selectivity) can be compensated by having a longer $t_{\pi}$, and vice-versa.}
If we rescale the detuning  $\nu \equiv \Delta t_{\pi}$ (simply a scaling in the function with respect to $\Delta$ by $t_{\pi}$; the shape, and hence, the selectivity is the same), $\eta =(\nu-n \chi t_{\pi})/\pi$.
Consequently, it is clear that Eq.~(\ref{EQ_ENSpe}), up to a scaling, is determined simply by $\chi t_{\pi}$.
For instance, this means that having larger $\chi$ (more selective) is equivalent to having the duration of the protocol $t_{\pi}$ longer. Also, a low $\chi$ (less selective) can be compensated by a longer $t_{\pi}$.

\subsection{Parity mapping}
\label{apx_sub:parity-mapping}
In this section, we provide the analogous expressions for the qubit probability for the case of mapping the parity of the cavity state. We show that this mapping is inaccurate, which introduces a scaling and offset corrections to the ideal parity.

Parity mapping is done via a standard Ramsey spectrocopy sequence ($\pi/2$ pulse~-~wait~-~$\pi/2$ pulse). The ideal evolutions during the $\pi/2$ pulses and the waiting time are governed by the Hamiltonians $\hat H_p = (\Omega/2)\hat \sigma_y$  applied for a time $t_{\pi/2}=\pi/(2\Omega)$  and $\hat H_d=-\chi |e\rangle \langle e| \otimes \hat n$  applied for a time $t_w = \pi/\chi$, respectively.
Considering the qubit starting in $|g\rangle$ and the cavity in an arbitrary state $\rho$, both $\pi/2$ pulses realize a $\pi/2$ rotation on the qubit (along the $y$-axis on the Bloch sphere) with a conditional phase gate $\hat C_{\pi} = |g\rangle \langle g|\otimes \openone + |e\rangle \langle e|\otimes \hat P$, with $\hat P = e^{i\pi \hat n}$, implemented inbetween. This results in an ideal mapping of the parity onto the qubit, the probability of finding the qubit in the excited state being
\begin{equation}\label{EQ_Pid}
    p_e = \frac{1}{2}(1+P_{\text{id}}),
\end{equation}
where $P_{\text{id}}\equiv \text{Tr}(\rho \hat P)$ denotes the parity of the cavity state $\rho$. This can also be expressed as $P_{\text{id}}=-\langle \hat \sigma_z \rangle$, since even (odd) parity states are mapped to the south (north) pole of the Bloch sphere of the qubit.

\begin{figure}[!]
    \centering
    \includegraphics{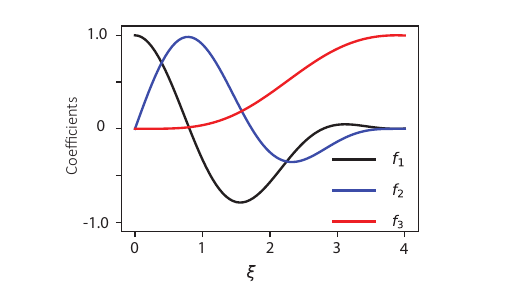}
    \caption{{\bf Parity mapping coefficients.} The coefficients for the parity computed using Eq.~(\ref{EQ_coeffs}) as a function of $\xi=\chi n/\Omega$.
    }
    \label{fig:coeffs}
\end{figure}

In a real scenario, however, the always-on dispersive coupling is also present during the $\pi/2$ pulses, which results in an actual Hamiltonian $\hat H_{pd}=(\Omega/2) \hat \sigma_y-\chi |e\rangle \langle e| \otimes \hat n$, which introduces a coherent error for the parity mapping.
A standard technique to partially counter this is to reduce the duration during the wait $t_w<\pi/\chi$. For simplicity, considering the initial cavity state to be a pure Fock state $\rho=|n\rangle \langle n|$, the actual dynamics of the Ramsey spectrocopy yield a parity mapping
\begin{align}
P^{\prime}&\equiv-\langle \hat \sigma_z \rangle \nonumber \\
&=f_1(\xi)\cos(\chi n t_w)-f_2(\xi)\sin(\chi n t_w)-f_3(\xi),
\end{align}
where $\xi \equiv \chi n/\Omega$ and 
\begin{widetext}
    \begin{eqnarray} \label{EQ_coeffs}
    f_1 &=&\frac{1}{(1+\xi^2)^2} \left( \sin^2(\frac{\pi}{2}\sqrt{1+\xi^2})+2\xi^2\cos(\frac{\pi}{2}\sqrt{1+\xi^2})(1-\cos(\frac{\pi}{2}\sqrt{1+\xi^2}))\right),\nonumber \\
    f_2 &=&\frac{2\xi}{(1+\xi^2)^{3/2}} \sin(\frac{\pi}{2}\sqrt{1+\xi^2})(1-\cos(\frac{\pi}{2}\sqrt{1+\xi^2})),\nonumber \\
    f_3 &=&\frac{\left(\xi^2+\cos(\frac{\pi}{2}\sqrt{1+\xi^2}) \right)^2}{(1+\xi^2)^2},
    \end{eqnarray}
\end{widetext}
for a general waiting time $t_w$. In the limit of infinitely-short $\pi/2$ pulses ($\xi \rightarrow 0$) and $t_w=\pi/\chi$, this equation reduced to the ideal parity $P^{\prime}=\cos(n\pi)=(-1)^n = P_{\text{id}}$.
The $f_1$ term contains most of the ideal parity contrast, the $f_2$ term results in a correction from the optimal waiting time $t_w<\pi/\chi$, and the $f_3$ term yields a positive offset to the parity (regardless of its sign). To illustrate the values of $f_1$, $f_2$, and $f_3$ for higher Fock states, we plot these coefficients against $\xi$ in Fig.~\ref{fig:coeffs}.
%For a larger Fock state, the offset term dominates and the parity tends to $P^\prime \rightarrow -1$.
By identifying the roles of these coefficients, we can conveniently define the parity errors
\begin{equation}
    P^{\prime} = P_{\text{id}}\:\eta -\zeta,
    \label{eqn:scaling_offset_error}
\end{equation}
where $\eta$ is a scaling error that incorporates the $f_1$ and $f_2$ terms, and $\zeta$ is an offset error coming from $f_3$.

By reversing the second $\pi/2$ pulse in the sequence, i.e., using $\hat H_{pd}=-(\Omega/2) \hat \sigma_y-\chi |e\rangle \langle e| \otimes \hat n$, the parity now reads
\begin{align}
P^{\prime}_{\text{rev}}&\equiv-\langle \hat \sigma_z \rangle \nonumber \\
&=-f_1(\xi)\cos(\chi n t_w)+f_2(\xi)\sin(\chi n t_w)-f_3(\xi).\nonumber \\
\end{align}
By using both $P^{\prime}$ and $P_{\text{rev}}^{\prime}$, we can correct the offset error by computing
\begin{eqnarray}
    P^{\prime}_{\text{cor}} &\equiv &\frac{P^{\prime}-P^{\prime}_{\text{rev}}}{2}\nonumber \\
    &=&f_1(\xi)\cos(\chi n t_w)-f_2(\xi)\sin(\chi n t_w)\nonumber \\
    &=&P_{\text{id}}\:\eta.
\end{eqnarray}
However, we can see that the scaling error remains.

\begin{figure}[!]
    \centering
    \includegraphics{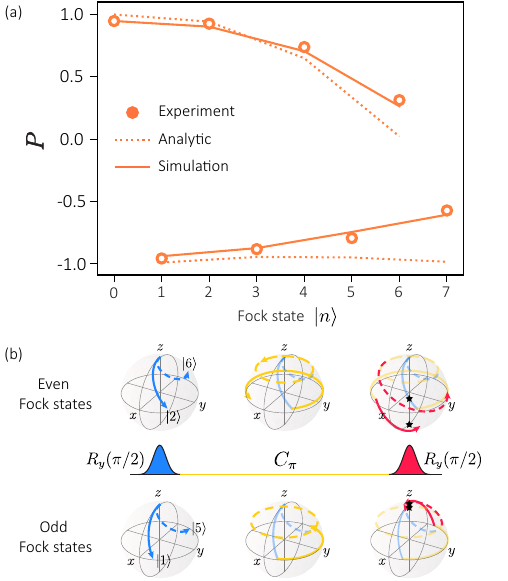}
    \caption{\textbf{Parity coherent errors.}(a) Measurement outcomes of parity $P$ under scaling and offset errors as a function of the Fock state $|n\rangle$ prepared in the cavity. They show good agreement with analytical, Eq.~(\ref{eqn:scaling_offset_error}), and simulated trends based on real device parameters. (b) Bloch spheres of the qubit state at each step of the parity mapping for even and odd Fock states. The mapping of the even states suffers more due to the offset error. Black stars represent the measured outcomes (projections on the $z$-axis).
    }
    \label{fig:coherent_error_exp}
\end{figure}

We experimentally observe $P'$ and the impact of the scaling and offset errors by measuring the parity of a series of Fock states with the Ramsey protocol, using 16~ns $\pi/2$ pulses and a 284~ns waiting time, see Fig.~\ref{fig:coherent_error_exp}a. Scaling error shows up as a decrease in overall contrast, and offset error appears as the skew between the even and odd Fock states. Fig.~\ref{fig:coherent_error_exp}b is an illustration of how the measurements of even Fock states are degraded more significantly than odd ones as $n$ increases. This is due to the tilted rotation axis asymmetrically impacting the Ramsey evolutions of the evens and the odds. If we were to flip the mapping of the evens and odds to the qubit measurement outcomes, then odd Fock states would degrade more significantly.

%%%%%%%%%%%%%%%%%%%%%%%%%%%%%%%%%%%%%%%%%%%%%%%%%%%%

%%%%%%%%%%%%%%%%%%%%%%%%%%%%%%%%%%%%%%%%%%%%%%%%%%%%
\section{Analysis of incoherent errors in observables}
\label{apx:incoherent-error}
\subsection{Qubit dephasing $T_{\phi}$}
\label{apx_sub:qubit-decoherence}
The primary limitation of excitation number sampling is qubit decoherence. This stems from the frequency-selectivity imposed by the finite dispersive coupling $\chi$. To be precise, coherence time imposes a limit on the maximum $\pi$-pulse duration before the qubit decoheres. The maximum pulse duration sets the frequency bandwidth, and thus selectivity, of the pulse. Because the qubit frequency shifts by $\chi n$ for each excitation $n$ of the cavity, the maximum selectivity of the pulse then sets the minimum dispersive frequency shift $\chi$ necessary to resolve the excitation number.
For the qubit, there are two loss channels to consider: energy decay and dephasing, which are characterized by their respective coherence times $T_1$ and $T_{\phi}$.
While standard cQED setups can reliably achieve $T_1$s in the range of several tens to hundreds of microseconds~\cite{kjaergaard_superconducting_2020}, ensuring a consistent $T_{\phi}$ proves to be a challenging task~\cite{gargiulo_fast_2021}.

In what follows, we will analyze excitation number and parity mapping only under qubit dephasing. For the case of excitation number, Eq.~(\ref{EQ_qstateENS}) in Appendix~\ref{apx_sub:ens-mapping-selectivity} can be adapted to account for qubit dephasing 
by splitting the dynamics into small time intervals $\Delta t=t_{\pi}/N$ with $N\rightarrow \infty$ in which we repeatedly apply the loss channel $\varepsilon(\cdot)\equiv \sum_j E_j \cdot E_j^{\dagger}$, where $E_j$ are the Kraus operators, and the unitary evolution $\hat U_{\Delta t}\equiv e^{-i\hat H \Delta t}$,

\begin{widetext}
\begin{align}\label{EQ_qstateENSTphi}
    \rho_q(t_{\pi})
    &=\sum_n \langle n| \cdots \hat U_{\Delta t}\varepsilon(\hat U_{\Delta t}\varepsilon (|g\rangle \langle g |\otimes \rho) \hat U_{\Delta t}^{\dagger}) \hat U_{\Delta t}^{\dagger}\cdots|n\rangle\nonumber \\
    &=\sum_n \langle n|\rho |n\rangle \left(\cdots \hat U_{n,\Delta t}\varepsilon(\hat U_{n,\Delta t}\varepsilon (|g\rangle \langle g |) \hat U_{n,\Delta t}^{\dagger}) \hat U_{n,\Delta t}^{\dagger}\cdots\right),\nonumber \\
    p_e^{\prime}(t_{\pi}) 
    &= \sum_n \rho_{nn} \langle e| \left(\cdots \hat U_{n,\Delta t}\varepsilon(\hat U_{n,\Delta t}\varepsilon (|g\rangle \langle g |) \hat U_{n,\Delta t}^{\dagger}) \hat U_{n,\Delta t}^{\dagger}\cdots\right)|e\rangle.
\end{align}
\end{widetext}

Note that the error channel ($\varepsilon$) only acts on the qubit, while the Hamiltonian of $\hat U_{\Delta t}$ act on the Fock state $|n\rangle$ of the cavity. This allows for the simplification, where a specific $\hat U_{n,\Delta t}\equiv e^{-i\hat H_n\Delta t}$ with $\hat H_n=\Delta_n |e\rangle \langle e|+(\Omega/2)\hat \sigma_y$ only acts on the qubit. The third line then presents the qubit probability and cavity excitation mapping, analogous to Eq.~\ref{EQ_ENSpe} in Appendix~\ref{apx_sub:ens-mapping-selectivity}. This last expression can be written as $p^{\prime}_e(t_{\pi})=\sum_n \rho_{nn} w_n$, where the weight $w_n$ is the probability of the qubit being in excited state after starting in $|g\rangle$.% and undergoing the dynamics acording to $\hat H_n$ under dephasing loss. 
With the same selectivity assumption ($\chi/\Omega \gg 1$) used in Appendix~\ref{apx_sub:ens-mapping-selectivity}, we arrive at 
\begin{equation}
    p^{\prime}_e(\Delta_n=0)\approx \rho_{nn} w,
\end{equation}
where the weight $w$ is obtained in the same way as $w_n$ but with a Hamiltonian $\hat H=(\Omega/2)\hat\sigma_y$ under dephasing. In this way, qubit dephasing scales the probability of all excitation number $\rho_{nn}$ with the same magnitude. The weight $w$ can be computed by solving the qubit dynamics under decoherence with the Lindblad master equation 
\begin{equation}
    \dot \rho = -i[H,\rho]+\hat J\rho \hat J^{\dagger}-\frac{1}{2}\{\rho,\hat J^{\dagger}\hat J\},
\end{equation}
where $\hat J\equiv \sqrt{2/T_{\phi}}|e\rangle \langle e|$ is the jump operator for qubit dephasing, which translates to solving a second-order differential equation. After a tedious but straightforward calculations, we obtain a weight
\begin{widetext}
    \begin{eqnarray}
    w&=&\frac{1}{2}(1-e^{-\gamma\pi}(\cos(\sqrt{1-\gamma^2}\pi)+\frac{\gamma}{\sqrt{1-\gamma^2}}\sin(\sqrt{1-\gamma^2}\pi)), \:\: \text{for}\: \gamma<1, \nonumber \\
    w &=& \frac{1}{2}(1-e^{-\gamma\pi}(\frac{\gamma+\sqrt{\gamma^2-1}}{2\sqrt{\gamma^2-1}}e^{\sqrt{\gamma^2-1}\pi}+\frac{-\gamma+\sqrt{\gamma^2-1}}{2\sqrt{\gamma^2-1}}e^{-\sqrt{\gamma^2-1}\pi})), \:\: \text{for}\: \gamma>1,
\end{eqnarray}
\end{widetext}
where $\gamma \equiv 1/(2T_{\phi}\Omega)$ and the first (second) line represents the small (over) dephasing case.
In most cases, we have small dephasing such that (up to the second order in $\gamma$) the weight can be approximated as $w\approx (1+e^{-\gamma\pi})/2$, and the mapping
\begin{equation}
    p^{\prime}_n\equiv p^{\prime}_e(\Delta_n=0) \approx \rho_{nn} \times \frac{1}{2}(1+e^{-\frac{t_{\pi}}{2T_{\phi}}}).
\end{equation}

For the case of parity mapping, we simplify the calculation by assuming that dephasing is only present during the waiting time, since the $\pi/2$ pulses are much shorter in time.
Hence, only the off-diagonal elements of the qubit density matrix are degraded by a factor $e^{-t_w/T_{\phi}}$. The parity mapping consequenctly yields
\begin{equation}
    P^{\prime} \equiv 2p^{\prime}_e-1=P_{\text{id}}\times e^{-\frac{t_w}{T_{\phi}}}.
\end{equation}

\subsection{Qubit thermal population}
\label{apx_sub:thermal-population}

Both excitation number and parity mapping rely on the qubit initialized in the ground state $|g\rangle$. In reality, the qubit might be in a mixed state with some probability being in the excited state before the mapping protocol is performed. Experimentally, the residual excited probability is mainly due to imperfect state preparation from the GRAPE pulses.
To analyze the effect of this, we assume an initial state $\rho_{qth}\otimes \rho$, where $\rho_{qth}=(1-\lambda)|g\rangle \langle g|+\lambda |e\rangle \langle e|$, with $\lambda$ being the probability of the qubit in the excited state.

For excitation number mapping, following the derivation in Eq.~(\ref{EQ_qstateENS}), results in 
% \begin{widetext}
\begin{eqnarray}\label{EQ_qstateENSthermal}
    &&p^{\prime}_e(t) =\nonumber \\
    &&\sum_n \rho_{nn} \langle e|e^{-i\hat H_n t}((1-\lambda)|g\rangle \langle g|+\lambda |e\rangle \langle e|)\otimes \rho e^{i\hat H_nt}|e\rangle.\nonumber \\
\end{eqnarray}
% \end{widetext}
When evaluating the term $\langle e| \cdots |e\rangle$ in the summation at $t_{\pi}=\pi/\Omega$, we note two cases: (i) $\Delta_n=0$ and (ii) $\Delta_n\gg \Omega$.
For (i), the unitary is $e^{-i\hat H_n t_{\pi}}=e^{-i\hat \sigma_y \pi/2}$, which is a $\pi$ rotation of the qubit. In this case, we have $\langle e| \cdots |e\rangle=1-\lambda$.
For (ii), the unitary is $e^{-i\hat H_n t_{\pi}}\approx e^{-i\Delta_n t |e\rangle \langle e|}$. In this case, we have $\langle e| \cdots |e\rangle=\lambda$.
Both (i) and (ii) are essential for excitation number mapping:
\begin{equation}
    p_n^{\prime}\equiv p^{\prime}_e(\Delta_n=0)=(1-\lambda)\rho_{nn}+\lambda \sum_{j\ne n}\rho_{jj},
\end{equation}
where the biggest contribution comes from $\rho_{nn}$ with weight $1-\lambda$ accompanied by small contributions from other excitation probabilities $\rho_{jj}$ ($j\ne n$) each with weight $\lambda$ as they are far away from $\Delta_n=0$ (we assume $\Delta_n\gg \Omega$).
By noting that $\sum_{j\ne n} \rho_{jj}=(1-\rho_{nn})$ ($\text{Tr}(\rho)=1$), we have 
\begin{equation}
    p_n^{\prime}=\rho_{nn}(1-2\lambda)+\lambda,
\end{equation}
where we see that the thermal population of the qubit introduces a scaling and offset error to the ideal excitation number.
However, if $\lambda$ is known, the excitation number measurement can be corrected as $\rho_{nn}=(p_n^{\prime}-\lambda)/(1-2\lambda)$, where $p_n^{\prime}$ is the measured value.

For parity mapping, solving the dynamics starting with the qubit in a mixed state can be done by separating the two possible initial states. 
When the qubit starts in $|g\rangle$ state, with a probability $1-\lambda$, 
the parity is given by Eq.~(\ref{EQ_Pid}). On the other hand, when the qubit starts in $|e\rangle$ state, with a probability $\lambda$, we analogously arrive at $p_e = (1-P_{\text{id}})/2$. Combining both contributions,
\begin{equation}
    p^{\prime}_e = (1-\lambda)\frac{1+P_{\text{id}}}{2}+\lambda\frac{1-P_{\text{id}}}{2},
\end{equation}
which leads to a parity 
\begin{equation}
    P^{\prime}\equiv 2p^{\prime}_e-1=(1-2\lambda)P_{\text{id}}.
\end{equation}
The thermal population of the qubit only results in a scaling error that can be corrected as $P_{\text{id}}=P^{\prime}/(1-2\lambda)$, where $P^{\prime}$ is the measured value and $\lambda$ has been previously characterized.

%%%%%%%%%%%%%%%%%%%%%%%%%%%%%%%%%%%%%%%%%%%%%%%%%%%%
\section{Estimator for $\rho$}
\label{apx: estimation}
\subsection{Linear inversion}
\label{apx_sub:linear-inversion}

The oldest and simplest procedure to build an estimator for $\rho$ is called linear inversion. This method consists of interpreting the relative frequencies of measurement outcomes as probabilities and then inverting Born's rule through a least-squares (LS) inversion to obtain a $\rho_\text{LS}$ that predicts these probabilities. 

Born's rule relates the outcome probability $p_k$ of a certain measurement observable $\hat E_k$ to $\rho$ as
\begin{equation}
    p_k = \Tr (\rho \hat E_k).
\end{equation} 
Upon many measurement repetitions, we build a histogram and approximate each $p_k$ with the corresponding relative frequency of the outcome $k$. For ORENS, each measurement observable is defined by a displacement $\hat D(\alpha_k)$ and an excitation number $n$, and can be writen as 
\begin{equation}\label{EQ_obsdis}
    \hat E_{n,\alpha_k } = \hat D_{\alpha_k}|n \rangle \langle n | \hat D_{-\alpha_k},
\end{equation}
with the corresponding probability $p_n = \text{Tr}(\hat E_{n, \alpha_k} \rho)$. Let us define a $(D^2-1) \times D^2$ measurement matrix $M$ to describe the set of ORENS measurements as
\begin{equation}
    M = 
    {\begin{pmatrix}
        {\vec {E}}_{n, \alpha_1}\\
        {\vec {E}}_{n, \alpha_2}\\
        \vdots \\
        {\vec {E}}_{n, \alpha_{D^2 - 1}}
    \end{pmatrix}\\},\label{EQ_Mmatrix}
\end{equation} where $D$ is the cut-off dimension of the cavity state and $\vec E$ is the row-wise vectorized form of $\hat E$ (truncated to $D\times D$). Vectorizing $\rho$ column-wise to get $\vec \rho$ of length $D^2$ and writing the outcome probabilities as a vector of length $D^2-1$, $\vec p$, we can then write the matrix equation
\begin{equation}
    M \vec \rho = \vec p.
\label{eqn:linear-equation}
\end{equation}

Linear inversion corresponds to inverting this system using the observed relative frequencies $\vec p$ to derive $\vec \rho$. Because $M$ is not a square matrix, the system is solved using the Moore-Penrose pseudoinverse as 
\begin{equation}
    \vec \rho = (M^\dagger M)^{-1} M^\dagger \vec p = M^+ \vec p.
\end{equation}
% For $M^\dagger M$ to be invertible, the  $D^2-1$ measurement observables must be independent i.e. the set of measurements must be informationally complete. 
However, $M^\dagger M$ is not invertible since $\text{det}(M^\dagger M)=0$ as $M$ has fewer rows than columns~\cite{krisnanda2023tomographic}. Consequently, in our work, we parameterize and vectorize $\rho$ using the real and imaginary components of the upper triangular off-diagonal elements of $\rho$, as well as the diagonal elements except for the last one. This parametrization yields a vector $\vec Y$ of length $D^2 - 1$, which completely characterizes the state. 

Different parameterizations of $\rho$ are related linearly, i.e., $\vec \rho = K \vec Y + \Theta$. Using Eq.~(\ref{eqn:linear-equation}), and defining 
\begin{eqnarray}
    \mathcal{M} &\equiv& MK,\nonumber \\
     V &\equiv& M\Theta,\label{EQ_transform}
\end{eqnarray}
we obtain the modified linear equation and its inverse (linear inversion) 
\begin{eqnarray}\label{EQ_obsinputlin}
     \vec p &=& \mathcal{M}\vec Y + V,\nonumber \\
     \vec Y_{est} &=&\mathcal{M}^{-1} (\vec p-V),
\end{eqnarray} 
where the matrix $\mathcal{M}$ is now a square matrix. The density matrix corresponding to $\vec Y_{est}$ is the least-squares estimate $\rho_{LS}$.

\subsection{Optimizing the set of measurements}
\label{apx_sub:optimizing-set-measurements}

% Feasible reconstruction of a CV system requires defining a truncation dimension \(D\), within which all states of interest are contained. 

For a given truncation dimension \(D\), we then perform a classical optimization algorithm to obtain the set of measurement observables. The optimization involves determining the set of displacement points \(\{\alpha_k\}_{k=1}^{D^2-1}\) and the corresponding excitation number \(n\) to measure. This set enables the reconstruction of \emph{any} state that can be truncated within dimension \(D\), without any further assumption.

The optimization is executed using a gradient-descent method, where the cost function is the condition number (CN) of the matrix $\mathcal{M}$ in Eq.~(\ref{EQ_obsinputlin}), defined as 
\begin{equation}\label{EQ_CN}
    ||\mathcal{M}|| \:\:||\mathcal{M}^{-1}||, 
\end{equation}
where $||.||$ denotes the Euclidean norm. The CN quantifies the worst-case error amplification from the observables to the estimated state, i.e., when solving for $\vec Y_{est}$ in Eq.~(\ref{EQ_obsinputlin}) given the observables $\vec p$~\cite{bhatia_matrix_1997}. With Eqs.~(\ref{EQ_obsdis}), (\ref{EQ_Mmatrix}), and (\ref{EQ_transform}), the CN can be computed numerically given the choice of photon number $n$ for measurement and the set of displacement points.

Our gradient-descent algorithm begins with randomly-selected initial displacement points and a fixed photon number $n$ for the excitation measurement. The algorithm iteratively updates the displacement points based on the gradient of the cost function, ultimately providing an optimized set of displacement points $\{\alpha_k\}$ corresponding to the minimum CN, for a given $n$. We repeat this gradient descent for all photon numbers $n \in[0,D-1]$ to find the $n$ that yields the smallest CN. We consistently find that, for a given dimension $D$, the optimal CN is obtained for $n=D-1$. Lastly, we examined the scenario in which each of the $D^2-1$ displacements employs a distinct excitation number measurement $n$. However, this approach does not significantly reduce the CN, although it does introduce increased experimental complexity. The code for deriving the set of optimized displacements is available on GitHub.

\begin{figure}[h]
    \centering
    \includegraphics{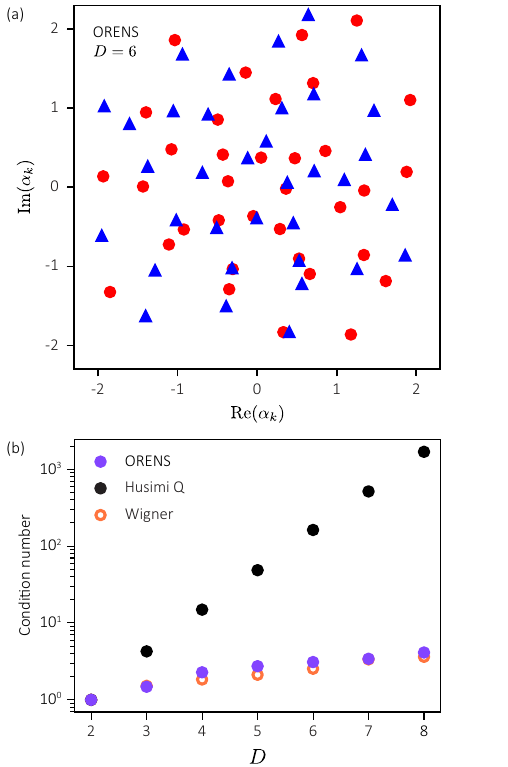}
    \caption{(a) Optimized displacement points of ORENS for a given truncation dimension $D=6$, all with excitation number $n=5$ measurement. Two different exemplary sets (blue triangles and red circles) are displayed. Both of these sets give condition numbers $\approx 3.1$, and comparable reconstruction fidelities. (b) Optimized condition number of ORENS, Wigner, and Husimi Q against truncation dimension of $\rho$.}
    \label{fig: CN_protocols}
\end{figure}

The sets of displacement points that yield a low CN are not unique, as exemplified in Fig.~\ref{fig: CN_protocols}a, where we show two different sets of optimized displacement points for ORENS given $D=6$ with a CN of $\approx 3.1$. Furthermore, when used for state reconstruction, simulations (similar to the one in Fig.~\ref{fig: Fig4}a) show that both sets offer an average fidelity $>0.98$. The fact that the displacement sets are not unique facilitates the incorporation of hardware constraints into the optimization algorithm. For instance, to mitigate readout distortion resulting from the residual cross-Kerr interaction when the cavity contains a high photon number, a constraint of $|\alpha_k|\le 2$ can be readily integrated into the optimization.

For benchmarking purposes, we also obtained the optimized displacement points for the Husimi-Q and the Wigner functions. The Husimi-Q function corresponds to ORENS in the case where the excitation number is constrained to measuring only zero excitations, $n=0$. For the Wigner function, the photon number projection operator $|n\rangle \langle n|$ in Eq.~(\ref{EQ_obsdis}) is simply replaced by the parity operator $\hat P$.

% The measurement observables for ORENS are optimized using a gradient-descent method over the displacements and excitation number $n$ to minimize the condition number (CN) of $W$, where the condition number captures the degree of error amplification. 
%An example code will be made available on GitHub soon.

The CN achieved by ORENS across dimensions is comparable to that of Wigner, indicating they have near-equivalent theoretical reconstruction capabilities (see Fig.~\ref{fig: CN_protocols}b). However, the optimal CN for Husimi Q-function scales unfavorably beyond $D>2$, illustrating that robust reconstruction is infeasible for large states. 

% Taking a closer look, it is observed that the Wigner function has slightly lower CN for a few of the dimensions. 

% \begin{table}
%     \centering
%     \begin{tabular}{ccc}
%         D & ORENS & Wigner\\
%         2 &  1.000& 1.000\\
%         3 &  2.482& 1.524\\
%         4 &  2.283& 1.838\\
%         5 &  2.736& 2.118\\
%         6 &  3.984& 2.565\\
%     \end{tabular}
%     \caption{Caption.}
%     \label{tab:condition_number}
% \end{table}
\subsection{Bayesian inference}
\label{apx_sub:bayesian}

Accurately inferring the quantum state of a system from measurement outcomes is a crucial task in quantum state reconstruction. In this section, we will motivate the use of Bayesian inference to process measurement outcomes and build the optimal estimator for $\rho$. For a deeper analysis, please refer to~\cite{blume-kohout_optimal_2010}, and for details on the specific methodology used in our work, please refer to~\cite{lukens_practical_2020}. 

The most notable limitation of linear inversion is that the estimated $\rho_\text{LS}$ frequently has negative eigenvalues, indicating that it cannot represent a physical state. Maximum likelihood estimation (MLE) was adopted as a convenient way to impose physicality on $\rho_\text{LS}$, and has been the dominant approach to quantum state reconstruction in recent years. Intuitively, it returns a single non-negative state $\rho_\text{MLE}$ that fits the observed data $\mathcal{D}$ as precisely as possible by maximizing the likelihood function, 
\begin{equation}
    \rho_\text{MLE} = \text{arg} \max_\rho L_\mathcal{D}(\rho),
\end{equation}
where $L_\mathcal{D}(\rho) \propto p(\mathcal{D}|\rho)$. An efficient way to obtain $\rho_\text{MLE}$ is described in Ref.~\cite{smolin_efficient_2012}. However, the MLE method does not quantify the level of uncertainty of the result, and most critically, $\rho_\text{MLE}$ often has zero eigenvalues. Consequently, it predicts exactly zero probability for every measurement outcome $|\psi\rangle \langle \psi |$ such that $\langle \psi | \rho | \psi \rangle = 0$. This implication of absolute certainty that a certain outcome will not be observed cannot reasonably be justified by a finite amount of data. The underlying flaw is that maximizing the likelihood is frequentist by nature; it interprets the observed relative frequencies of the measurement outcomes as probabilities, and then seeks to fit the probabilities as precisely as possible. However, the goal of state estimation extends beyond explaining the data to predicting future evolutions and states. Thus, estimation should involve the knowledge of the system being estimated, especially its uncertainty. 

In our work, we employed a Bayesian inference technique stemming from a different perspective on statistics that 1) considers many of the possible $\rho$, 2) accounts for experimental uncertainty explicitly through Bayes' rule, and 3) guarantees the most accurate estimate of the true $\rho$ that can be made from the data~\cite{blume-kohout_optimal_2010, williams_quantum_2017, granade_practical_2016}. Parameterizing $\rho(\bold x)$ by some vector $\bold x$, such that any value of $\bold x$ within its support returns a physical $\rho$, Bayes' theorem states that posterior probability distribution of $\bold x$ follows as

\begin{equation}
    \pi(\bold x) = \frac{1}{Z}L_{\mathcal{D}}(\bold x) \pi_0 (\bold x),
\end{equation} where $L_{\mathcal{D}}(\bold x)$ is the same likelihood as in MLE, $\pi_0 (\bold x)$ is the prior distribution that encapsulates any knowledge or beliefs about $\rho$ before the experiment, and $Z$ a normalizing constant. This posterior distribution gives us access to the expectation value of any function $\phi$ of $\rho$ via
\begin{equation}
    \langle \phi(\rho) \rangle = \int d \bold x \pi(\bold x) \phi (\rho(\bold x)).
\end{equation} 
Evaluating integrals of this form is numerically challenging due to the high dimensionality and complicated features. We overcome this challenge by employing the efficient Bayesian inference strategy~\cite{lukens_practical_2020} that is computationally practical and straightforward to implement through a combination of well-chosen parameterization of $\rho$ and likelihood, and the Markov Chain Monte Carlo (MCMC) sampling algorithm. Intuitively, the algorithm draws random samples of possible $\rho$ from a distribution across all physical states. These states are weighted by a pseudo-likelihood function that scales inversely with the distance between the sample and $\rho_\text{LS}$. These samples allow us to estimate any function of $\rho$ via
\begin{equation}
    \langle \phi(\rho) \rangle \approx \frac{1}{R} \sum_{r=1}^R \phi (\rho_r)
\end{equation} where R is the total number of MCMC samples. In detail, we chose the following parameters for Bayesian inference: $\alpha=1$ for a uniform prior on all possible physical density matrices; $\sigma = 1/N, \text{ with } N = 1000\cdot(D^2-1)$, as the variance for the pseudo-likelihood function around $\rho_\text{LS}$; and $2^{10}$ MCMC samples with thinning parameter $2^7$ to reduce serial correlation in the chain.

All simulated and experimental fidelities in this work were calculated with the Bayesian mean estimator (BME), defined as
\begin{equation}
    \rho_\text{BME} = \frac{1}{R}\sum_{r=1}^R \rho_r,
\end{equation} which stands as the most accurate estimator of the true $\rho$. Error bars in the main text represent the standard deviation across $\rho_\text{BME}$ fidelities of all the set of reconstructed states (either Fock states or cat states). We demonstrate how the performance of BME surpasses that of MLE, as expected, in Tables~\ref{tab:fock_dimensions_BME} and \ref{tab:cat_t2_BME} for ORENS reconstruction both across dimensions and decoherence regimes, respectively.

\begin{table}[ht!]
    \centering
    \begin{tabular}{ P{1cm}|P{1cm}|P{1cm}| P{3cm} }
        $D$ & $\bar F_\text{BME}$ & $\bar F_\text{MLE}$ & $\Delta = \bar F_\text{BME} -\bar F_\text{MLE}$\\
        \hline
        2 & 0.992 & 0.987 & 0.005 \\
        3 & 0.988 & 0.979 & 0.009 \\
        4 & 0.973 & 0.958 & 0.015 \\
        5 & 0.950 & 0.933 & 0.017 \\
        6 & 0.939 & 0.918 & 0.021 \\
    \end{tabular}
    \caption{Average ORENS reconstruction fidelity using BME and MLE across all Fock state superpositions for a given cut-off dimension, as plotted in Fig.~\ref{fig: Fig4}a.}
    \label{tab:fock_dimensions_BME}
% \end{table}

% \begin{table}
    \centering
    \begin{tabular}{ P{1cm}|P{1cm}|P{1cm}| P{3cm} }
        $T_\phi (\mu\text{s})$ & $\bar F_\text{BME}$ & $\bar F_\text{MLE}$ & $\Delta = \bar F_\text{BME} -\bar F_\text{MLE}$\\
        \hline
        22.4 & 0.947 & 0.932 & 0.015 \\
        10.4 & 0.946 & 0.927 & 0.018 \\
        3.48 & 0.944 & 0.931 & 0.013 \\
        1.02 & 0.875 & 0.856 & 0.019 \\
        0.535 & 0.867 & 0.851 & 0.016 \\
    \end{tabular}
    \caption{Average ORENS reconstruction fidelity using BME and MLE across all four cat states for different engineered qubit $T_\phi$, as plotted in Fig.~\ref{fig: Fig4}b.}
    \label{tab:cat_t2_BME}
\end{table}

%\section{Estimation and simulation code}
%\label{apx: code}
%The code used to simulate ORENS with real hardware parameters and to process measurement outcomes and estimate $\rho$ will be available on GitHub soon. 

%\bibliographystyle{unsrt}
%\bibliographystyle{apsrev4-1}

%% remember to think about potential ref to Qin et al PRL (2021).
\clearpage
\bibliography{zotero-refs}    %use a bibtex bibliography file
%%%%%%%%%%%%%%%%%%%%%%%%%%%%%%%%%%%%%%%%%%%%%%%%%%%
% \bibliography{Adrian_references}
\end{document}